\documentclass[12pt]{article}
\usepackage{amsmath} 
\input{epsf} 
\newcommand\as{\alpha_{\mathrm{S}}} 
\newcommand\f[2]{\frac{#1}{#2}} 
\def\beq{\begin{equation}} 
\def\eeq{\end{equation}} 
\def\beeq{\begin{eqnarray}} 
\def\eeeq{\end{eqnarray}} 
 
\def\to{\rightarrow}
\def\ito{\leftarrow} 
\def\nn{\nonumber}

\def\ms{${\overline {\rm MS}}$}
\def\bqt{{\bf q_T}}

\def\ep{\epsilon}

\def\qi{\{q_i\}}

\begin{document} 
\begin{titlepage}
\begin{flushright}
ZU-TH 25/13\\
IFUM-1017-FT
\end{flushright}
\renewcommand{\thefootnote}{\fnsymbol{footnote}}
\vspace*{1cm}

\begin{center}
{\Large \bf Universality of transverse-momentum resummation\\[0.3cm]
and hard factors at the NNLO}
\end{center}

\par \vspace{2mm}
\begin{center}
{\bf Stefano Catani${}^{(a)},$ Leandro Cieri${}^{(b)},$
Daniel de Florian${}^{(c)}$, Giancarlo Ferrera${}^{(d)},$\\
}
and
{\bf Massimiliano Grazzini${}^{(e)}$\footnote{On leave of absence from INFN, Sezione di Firenze, Sesto Fiorentino, Florence, Italy.}}\\

\vspace{5mm}

$^{(a)}$ INFN, Sezione di Firenze and Dipartimento di Fisica e Astronomia,\\ 
Universit\`a di Firenze,
I-50019 Sesto Fiorentino, Florence, Italy\\

$^{(b)}$ Dipartimento di Fisica, Universit\`a di Roma ``La Sapienza'' and\\
INFN, Sezione di Roma, I-00185 Rome, Italy\\

${}^{(c)}$ Departamento de F\'\i sica, FCEYN, Universidad de Buenos Aires,\\
(1428) Pabell\'on 1 Ciudad Universitaria, Capital Federal, Argentina\\

${}^{(d)}$ Dipartimento di Fisica, Universit\`a di Milano and\\ INFN, Sezione di Milano,
I-20133 Milan, Italy\\

$^{(e)}$ Institut f\"ur Theoretische Physik, Universit\"at Z\"urich, CH-8057 Z\"urich, Switzerland

\vspace{5mm}

\end{center}

\par \vspace{2mm}
\begin{center} {\large \bf Abstract} \end{center}
\begin{quote}
\pretolerance 10000

We consider QCD radiative corrections to the production of
colourless high-mass systems in hadron collisions.
The logarithmically-enhanced
contributions at small transverse momentum are treated to all 
perturbative orders by a universal resummation formula that depends on a single
process-dependent hard factor.
We show that the 
hard factor is directly related to the
all-order virtual amplitude of the corresponding partonic process.
The direct relation is universal (process independent),
and it is expressed by an all-order factorization formula that we explicitly
evaluate up to the next-to-next-to-leading order (NNLO) 
in QCD perturbation theory. 
Once the NNLO scattering amplitude is available,
the corresponding hard factor
is directly determined:
it controls NNLO contributions 
in resummed calculations at full next-to-next-to-leading logarithmic accuracy,
and it can be used in applications of the $q_T$ subtraction formalism to perform 
fully-exclusive perturbative calculations up to NNLO.
The universality structure of the hard factor and its explicit NNLO form are also
extended to the related formalism of threshold resummation.

\end{quote}

\vspace*{\fill}
\begin{flushleft}
November 2013

\end{flushleft}
\end{titlepage}

\newpage
\setcounter{page}{1}
\setcounter{footnote}{1}
\renewcommand{\thefootnote}{\fnsymbol{footnote}}

\section{Introduction}
\label{sec:intro}

The transverse-momentum $(q_T)$
distribution of systems with high invariant mass $M$
produced in hadron collisions
is important for physics studies within and beyond the Standard Model (SM).
This paper is devoted to a theoretical study of QCD radiative corrections
to transverse-momentum distributions.

We consider the inclusive hard-scattering reaction
\begin{equation}
h_1(p_1)+h_2(p_2)\to F(\{q_i\})+X\, ,
\label{class}
\end{equation}
where the collision of the two hadrons $h_1$ and $h_2$ with momenta $p_1$ and
$p_2$ produces the triggered final state $F$, and $X$ denotes the accompanying
final-state radiation. 
The observed final state $F$ is a generic system of one or more
{\em colourless} particles,
such as lepton pairs (produced by Drell--Yan (DY) mechanism), photon pairs, 
vector bosons, Higgs boson(s), and so forth.
The momenta of these final state particles are denoted by $q_1$,$q_2$...$q_n$.
The system $F$ has
{\em total} invariant mass $M^2=(q_1+q_2+...q_n)^2$, transverse momentum $\bqt$
and rapidity $y$. We use $\sqrt{s}$ to denote the centre-of-mass energy 
of the colliding hadrons, which are treated in the massless
approximation ($s=(p_1+p_2)^2=2p_1\cdot p_2$).

The transverse-momentum cross section for the process in Eq.~(\ref{class})
is computable by using perturbative QCD.
However, in the small-$q_T$ region (roughly, in the region where
$q_T \ll M$) the convergence of the fixed-order perturbative expansion
in powers of the QCD coupling $\as$ is spoiled by 
the presence of large logarithmic 
terms of the type $\ln^n(M^2/q_T^2)$.
The predictivity of perturbative QCD can be recovered through the summation
of these logarithmically-enhanced contributions to all order in $\as$ 
\cite{Dokshitzer:hw,Parisi:1979se,Curci:1979bg,Collins:1981uk,Kodaira:1981nh,Collins:1984kg,Catani:vd, deFlorian:2000pr, Catani:2000vq,
Catani:2010pd}. As already stated, we shall limit ourselves to considering the
production of systems $F$ of non-strongly interacting particles.
The all-order analysis of the $q_T$ distribution of systems $F$ that involve 
coloured QCD partons has just started to be investigated, by considering 
\cite{Zhu:2012ts}
the specific case in which $F$ is formed by a $t{\bar t}$ pair.

In the case of a generic system $F$ of colourless particles, the large
logarithmic contributions to the $q_T$ cross section can be systematically
resummed to all perturbative orders, and 
the structure of the resummed calculation can be organized in a
{\em process-independent} form 
\cite{Collins:1981uk, Collins:1984kg, Catani:2000vq, Catani:2010pd}.
The all-order resummation formalism was first developed for the DY process
\cite{Collins:1984kg} (and the kinematically-related process of two-particle
correlations in $e^+e^-$ annihilation \cite{Collins:1981uk}). The
process-independent extension of the formalism has required 
two additional main steps:
the understanding of
the all-order process-independent structure of the Sudakov form factor
(through the factorization of a single process-dependent hard factor)
\cite{Catani:2000vq}, 
and the complete generalization to processes that are initiated by
the gluon fusion mechanism \cite{Catani:2010pd}.

The all-order process-independent form of the resummed calculation
has a factorized structure, whose resummation factors are 
(see Sect.~\ref{sec:resu}) the (quark and gluon) Sudakov form factor,
process-independent {\em collinear} factors and a process-dependent
{\em hard} or, more precisely (see Sect.~\ref{sec:hardvirtual}), hard-virtual  
factor. The resummation of the logarithmic contributions is controlled by these
factors or, equivalently, by a corresponding set of perturbative functions
whose perturbative {\em resummation coefficients} are computable order-by-order
in $\as$. The perturbative coefficients of the Sudakov form factor
are known, since some time
\cite{Kodaira:1981nh, Catani:vd, Davies:1984hs,deFlorian:2000pr,deFlorian:2001zd},
up to the second order in $\as$, and the third-order coefficient 
$A^{(3)}$ (which is necessary to explicitly perform resummation up to the
next-to-next-to-leading logarithmic (NNLL) accuracy) is also known 
\cite{Becher:2010tm}.
The next-to-next-to-leading order (NNLO) QCD calculation of the $q_T$ cross
section (in the small-$q_T$ region) has been explicitly
carried out in analytic form for two 
benchmark processes, namely, SM Higgs boson production \cite{Catani:2011kr}
and the DY process \cite{Catani:2012qa}. The results of 
Refs.~\cite{Catani:2011kr, Catani:2012qa} provide us with the
complete 
knowledge
of the process-independent {\em collinear} resummation coefficients up to the
second order in $\as$, and with the explicit expression of the hard coefficients
for these two specific processes. The purpose of the present paper (see below)
is to explicitly point out and derive the underlying {\em universal}
(process-independent) structure of the process-dependent {\em hard}
factor of the QCD all-order resummation formalism.

In Refs.~\cite{Gao:2005iu, Mantry:2009qz, Becher:2010tm,
GarciaEchevarria:2011rb, Chiu:2012ir, Collins:2012uy,
Becher:2012yn},
the resummation of small-$q_T$ logarithms has been reformulated 
in terms of factorization formulae that involve Soft Collinear Effective Theory
operators and (process-dependent) hard matching coefficients. 
The 
formulation of Ref.~\cite{Becher:2010tm}
has been applied \cite{Gehrmann:2012ze}
to the DY process by explicitly computing the (process-independent)
collinear quark--quark coefficients and the DY hard coefficient at the NNLO. The
results of this calculation \cite{Gehrmann:2012ze}
agree with those obtained in Ref.~\cite{Catani:2012qa}.
Transverse-momentum cross sections can also be studied by using other approaches
(which go beyond the customary QCD resummation formalism of the present paper)
that use transverse-momentum dependent (TMD) factorization
(see Refs.~\cite{Catani:1990xk, D'Alesio:2007jt, Rogers:2010dm, Collins:2011zzd} 
and references therein)
and, consequently, $k_{\perp}$-unintegrated parton densities and partonic cross
sections that are both TMD quantities.

In this paper we study the process-dependent hard factor of the transverse-momentum
resummation formula. We show that, for any process of the class in 
Eq.~(\ref{class}), the {\em all-order} hard factor has a universal structure that
involves a minimal amount of process-dependent information. The 
process-dependent information is entirely given by the scattering amplitude of the
Born-level partonic subprocess and its {\em virtual} radiative corrections.
Knowing the scattering amplitude, the hard-virtual resummation factor is determined
by a universal (process-independent) factorization formula. The universality
structure of the factorization formula has a {\em soft} (and collinear) origin, and
it is closely
(though indirectly) related to the universal structure of the infrared divergences
\cite{Catani:1998bh}
of the scattering amplitude.
This process-independent structure of the hard-virtual term, which generalizes the 
next-to-leading order (NLO) results of Ref.~\cite{deFlorian:2001zd}, is valid to
all perturbative orders, and we explicitly determine the 
process-independent form of the hard-virtual term up to the NNLO.
Using this general NNLO result, the  hard-virtual resummation factor for each
process of the class in Eq.~(\ref{class}) is straightforwardly computable up to its
NNLO, provided the corresponding scattering amplitude is known.

In the final part of the paper, we consider the related formalism of threshold
resummation \cite{Sterman:1986aj, Catani:1989ne} for the {\em total} cross section.
The process-independent formalism of threshold resummation also involves a
corresponding process-dependent hard factor. We shall show that this factor has a
universality structure that is analogous to the case of transverse-momentum
resummation. In particular, we directly relate the process-dependent hard factors
for transverse-momentum and threshold resummation in a form that is fully universal
and completely independent of each specific process (e.g., independent of the
corresponding scattering amplitude).

The knowledge of the NNLO hard-virtual term completes 
the $q_T$ resummation
formalism in explicit form up to full NNLL+NNLO accuracy.
This permits direct applications to NNLL+NNLO resummed calculations for any 
processes
of the class in Eq.~(\ref{class}) (provided the corresponding NNLO amplitude is
known), as already done for the cases of SM Higgs boson 
\cite{Bozzi:2005wk, deFlorian:2011xf, Wang:2012xs}
and DY \cite{Bozzi:2010xn,Guzzi:2013aja} production.

The NNLO information of the $q_T$ resummation formalism is also relevant in the
context of {\em fixed order} calculations. Indeed, it permits to carry out
fully-exclusive NNLO calculations by applying the $q_T$ {\em subtraction
formalism} of Ref.~\cite{Catani:2007vq}
(the subtraction counterterms of the formalism follow \cite{Catani:2007vq}
from the fixed-order expansion of the $q_T$ resummation formula, 
as in Sect.~2.4 of Ref.~\cite{Bozzi:2005wk}). 
The $q_T$ subtraction
formalism has been applied 
to the NNLO computation of Higgs boson \cite{Catani:2007vq,Grazzini:2008tf}
and vector boson production \cite{Catani:2009sm},
associated production of the Higgs boson with a $W$ boson \cite{Ferrera:2011bk}, 
diphoton production \cite{Catani:2011qz} 
and $Z\gamma$ production \cite{Grazzini:2013bna}.
The computations of 
Refs.~\cite{Catani:2007vq,Grazzini:2008tf,Catani:2009sm,Ferrera:2011bk}
were based on the specific calculation  of the NNLO hard-virtual coefficients
of the corresponding processes \cite{Catani:2011kr,Catani:2012qa}.
The computations of Refs.~\cite{Catani:2011qz, Grazzini:2013bna}
used the NNLO hard-virtual coefficients that are determined by applying
the universal form of the hard-virtual term that is derived and illustrated in the
present paper.

The paper is organized as follows. In Sect.~\ref{sec:resu} we recall 
the transverse-momentum resummation
formalism in impact parameter space, and we introduce our notation.
In Sect.~\ref{sec:procind} we present the explicit expressions of
the process-independent resummation coefficients up to NNLO.
Section~\ref{sec:hardvirtual} is devoted to the process-dependent hard
coefficients. We discuss and illustrate the universal all-order form of the 
hard-virtual coefficients by relating them to the process-dependent scattering
amplitudes, through the introduction of suitably subtracted hard-virtual 
matrix elements. The process-independent structure of the 
hard-virtual coefficients is explicitly computed up to the NNLO.
In Sect.~\ref{sec:thre} 
we extend our discussion and results on the universal structure of the
hard-virtual coefficients to the case of threshold resummation.
In Sect.~\ref{sec:summa} we summarize our results.
In Appendix \ref{appa} we report the explicit expressions of the NLO and NNLO
hard-virtual coefficients for DY, Higgs boson and diphoton production.


\section{Small-$q_T$ resummation}
\label{sec:resu}

In this section we briefly recall the
formalism of transverse-momentum resummation in impact parameter space 
\cite{Dokshitzer:hw, Parisi:1979se, Curci:1979bg, Collins:1981uk, 
Kodaira:1981nh, Collins:1984kg, Catani:vd, deFlorian:2000pr, Catani:2000vq,
Catani:2010pd}. We closely follow the notation of Ref.~\cite{Catani:2010pd}
(more details about our notation can be found therein).

We consider the inclusive-production process in Eq.~(\ref{class}), and we introduce the 
corresponding {\em fully} differential cross section
\begin{equation}
\label{diffxs}
\f{d\sigma_F}{d^2{\bqt} \;dM^2 \;dy \;d{\bf\Omega}} 
\,(p_1, p_2;\bqt,M,y,
{\bf\Omega} )
\;\;,
\end{equation}
which depends on the total momentum of the system $F$
(i.e. on the variables $\bqt, M, y$). The cross section
also depends on the set of additional variables
that controls the kinematics of the particles in the system $F$.
In Eq.~(\ref{diffxs}) these additional variables
are generically denoted as ${\bf\Omega}= \{\Omega_A,\Omega_B, \dots \}$
(correspondingly, we define $d{\bf\Omega} \equiv d\Omega_A d\Omega_B \dots)$. 
To be general, we do not explicitly specify these variables, and we only
{\em require} that the kinematical variables 
$\{\Omega_A,\Omega_B, \dots \}$ are {\em independent} of $\bqt, M$ and $y$ and
that the set of variables $\{\bqt, M, y, {\bf\Omega}\}$ {\em completely}
determines the kinematical configuration (i.e., the momenta $q_i$)
of the particles in the system $F$. For instance, if the system $F$ is formed by
two particles, there are only two variables in the set ${\bf\Omega}$, and they
can be the rapidity $y_i$ and the azimuthal angle $\phi(\bqt_i)$ of one of the
two particles (the particle with momentum $q_i$).
Note that the cross section in Eq.~(\ref{diffxs}) and the corresponding 
resummation formula can be straightforwardly integrated with respect to some of
the final-state variables $\{\Omega_A,\Omega_B, \dots \}$, thus leading to
results for observables that are more inclusive than the differential cross
section in Eq.~(\ref{diffxs}).

We also recall that we are considering the production of a system $F$ of
colourless particles or, more precisely, a system of non-strongly interacting
particles (i.e., $F$ cannot include QCD partons and their
fragmentation products).
Therefore, at the Born (lowest-order) level, the cross section in 
Eq.~(\ref{diffxs}) is controlled by the partonic subprocesses of quark--antiquark
($q{\bar q}$) annihilation,
\begin{equation}
q_f + {\bar q}_{f^\prime} \to F \;\;,
\label{qqpro}
\end{equation}
and gluon fusion,
\begin{equation}
g + g \to F \;\;.
\label{ggpro}
\end{equation}
Owing to colour conservation, no other partonic subprocesses can occur
at the Born level. 
More importantly (see below), the distinction between
$q{\bar q}$ annihilation and gluon fusion leads to relevant (and physical)
differences \cite{Nadolsky:2007ba, Catani:2010pd}
in the context of small-$q_T$ resummation.

To study the $\bqt$ dependence of the differential cross section
in Eq.~(\ref{diffxs}) within QCD perturbation theory, 
we introduce the following
decomposition:
\begin{equation}
\label{Fdec}
d\sigma_F =
d\sigma_F^{({\rm sing})} +
\; d\sigma_F^{({\rm reg})}
\;\;.
\end{equation}
Both terms in the right-hand side are obtained through convolutions of
partonic cross sections and the scale-dependent parton distributions 
$f_{a/h}(x,\mu^2)$  ($a=q_f, {\bar q}_f, g$ is the parton label) of the
colliding hadrons. We use parton densities as defined in 
the \ms\ factorization scheme, and $\as(q^2)$ is
the QCD running coupling in the \ms\ renormalization scheme. 
The partonic cross sections that enter the singular component (the first term
in the right-hand side of Eq.~(\ref{Fdec})) contain all 
the contributions that are enhanced
(or `singular') at small $q_T$. These contributions are proportional to
$\delta^{(2)}(\bqt)$ or to large logarithms of the type 
$\f{1}{q_T^2}\ln^m (M^2/q_T^2)$.
On the
contrary, the partonic cross sections of the second term in the right-hand side
of Eq.~(\ref{Fdec})
are regular (i.e. free of logarithmic terms)
order-by-order in perturbation theory as $q_T \to 0$. More precisely, the
integration of $d\sigma_F^{({\rm reg})}/d^2{\bqt}$ over the range 
$0 \leq q_T \leq Q_0$ leads to a finite result that, at each
fixed order in $\as$, {\em vanishes} in the limit $Q_0 \to 0$.
 
The regular component $d\sigma_F^{({\rm reg})}$
of the $q_T$ cross section
depends on the specific process in Eq.~(\ref{class}) that we are considering.
In the following we focus on
the singular component, $d\sigma_F^{({\rm sing})}$,
which has a universal all-order
structure. The corresponding resummation formula is written as \cite{Collins:1984kg,Catani:2000vq,Catani:2010pd}
\begin{align}
\label{qtycross}
&\f{d\sigma_F^{({\rm sing})}(p_1, p_2;\bqt,M,y,{\bf\Omega} )}{d^2{\bqt} \;dM^2 \;dy \;d{\bf\Omega}} 
=\f{M^2}{s}\sum_{c=q,{\bar q},g} \;
\left[d\sigma_{c{\bar c},F}^{(0)}\right]
\int \f{d^2{\bf b}}{(2\pi)^2} \;\, e^{i {\bf b}\cdot \bqt} \;
  S_c(M,b)\nn \\
& \;\;\;\; \times \;
\sum_{a_1,a_2} \,
\int_{x_1}^1 \f{dz_1}{z_1} \,\int_{x_2}^1 \f{dz_2}{z_2} 
\; \left[ H^F C_1 C_2 \right]_{c{\bar c};a_1a_2}
\;f_{a_1/h_1}(x_1/z_1,b_0^2/b^2)
\;f_{a_2/h_2}(x_2/z_2,b_0^2/b^2) \;
\;, 
\end{align}
where $b_0=2e^{-\gamma_E}$
($\gamma_E=0.5772\dots$ is the Euler number) is a numerical coefficient,
and the kinematical variables $x_1$ and $x_2$ are 
\begin{equation}
\label{xo}
x_1= \f{M}{\sqrt s} \;e^{+y} \;\;, \quad \quad
x_2=\f{M}{\sqrt s} \;e^{-y} \;\;.
\end{equation}
The right-hand side of Eq.~(\ref{qtycross}) involves the Fourier transformation
with respect to the
{\em impact parameter} ${\bf b}$ and two convolutions over the 
longitudinal-momentum fractions $z_1$ and $z_2$.
The parton densities 
$f_{a_i/h_i}(x,\mu^2)$ of the colliding hadrons are evaluated at the scale
$\mu= b_0/b$, which depends on the impact parameter.
The function $S_c(M,b)$ is the Sudakov form factor. This factor, which only
depends on the type ($c=q$ or $c=g$) of colliding partons, is universal
(process independent) \cite{Catani:2000vq}, and it resums the
logarithmically-enhanced contributions
of the form $\ln M^2b^2$ (the region $q_T \ll M$ corresponds to $Mb \gg 1$ in
impact parameter space). The all-order expression of $S_c(M,b)$ is \cite{Collins:1984kg}
\begin{equation}
\label{formfact}
S_c(M,b) = \exp \left\{ - \int_{b_0^2/b^2}^{M^2} \frac{dq^2}{q^2} 
\left[ A_c(\as(q^2)) \;\ln \frac{M^2}{q^2} + B_c(\as(q^2)) \right] \right\} 
\;\;,
\end{equation}
where $A_c(\as)$ and $B_c(\as)$ are perturbative series in $\as$,
\begin{equation}
A_c(\as)=\sum_{n=1}^\infty\left(\frac{\as}{\pi}\right)^n A^{(n)}_c~~~,~~~~~~~
B_c(\as)=\sum_{n=1}^\infty\left(\frac{\as}{\pi}\right)^n B^{(n)}_c\; .
\label{abfun}
\end{equation}
The perturbative coefficients $A^{(1)}_c, B^{(1)}_c, A^{(2)}_c$
\cite{Kodaira:1981nh, Catani:vd},  $B^{(2)}_c$ 
\cite{Davies:1984hs, deFlorian:2000pr, deFlorian:2001zd}
and $A^{(3)}_c$ \cite{Becher:2010tm} are explicitly known.

The factor that is symbolically denoted by
$\left[ d\sigma_{c{\bar c}, \,F}^{(0)} \right]$
in Eq.~(\ref{qtycross}) is the Born-level cross
section $d{\hat \sigma}^{(0)}$ (i.e., the cross section at its corresponding {\em lowest
order} in $\as$) of the partonic subprocesses $c{\bar c}\to F$ in 
Eqs.~(\ref{qqpro}) and (\ref{ggpro}) (in the case of the $c{\bar c}=q{\bar q}$
annihilation channel, the quark and antiquark can actually have different
flavours).
Making the symbolic notation explicit, we have
\begin{equation}
\label{sig0}
\left[ d\sigma_{c{\bar c}, \,F}^{(0)} \right]
= \f{d{\hat \sigma}_{c{\bar c}, \,F}^{(0)}}{M^2 \;d{\bf\Omega}} 
\,(x_1p_1, x_2p_2;
{\bf\Omega}; \as(M^2) )\;,
\end{equation}
where $x_1p_1^\mu$ ($x_2p_2^\mu$) is the momentum of the parton $c$ (${\bar c}$).
In Eq.~(\ref{qtycross}), we have included the contribution of both
the $q{\bar q}$ annihilation
channel ($c=q,{\bar q}$) and the gluon fusion 
channel ($c=g$); one of these two contributing channels may be absent 
(i.e. $\left[ d\sigma_{c{\bar c}, \,F}^{(0)} \right]=0$ in that channel), 
depending on the specific final-state system $F$.

The Born level factor $\left[ d\sigma_{c{\bar c}, \,F}^{(0)} \right]$
is obviously process dependent, although its process dependence is elementary
(it is simply due to the Born level scattering amplitude of the partonic process 
$c{\bar c}\to F$). The remaining process dependence of Eq.~(\ref{qtycross})
is embodied in the `hard-collinear' factor $\left[ H^F C_1 C_2 \right]$.
This factor includes a process-independent part and a process-dependent part.
The structure of the process-dependent part is the main subject of the 
present paper.


In the case of processes that are initiated at the Born level
by the $q{\bar q}$ annihilation
channel ($c=q$), the symbolic factor $\left[ H^F C_1 C_2 \right]$ in 
Eq.~(\ref{qtycross}) has the following explicit form \cite{Catani:2000vq}
\begin{align}
\label{what}
\left[ H^F C_1 C_2 \right]_{q{\bar q};a_1a_2}
 = H_q^F(x_1p_1, x_2p_2; {\bf\Omega}; \as(M^2))
\;\, C_{q \,a_1}(z_1;\as(b_0^2/b^2)) 
\;\, C_{{\bar q} \,a_2}(z_2;\as(b_0^2/b^2)) \;\;,
\end{align}
and the functions $H_q^F$ and $C_{q \,a}= C_{{\bar q} \,{\bar a}}$ have 
the perturbative expansion
\begin{align}
\label{hexp}
&H_q^F(x_1p_1, x_2p_2; {\bf\Omega}; \as) = 1+ \sum_{n=1}^\infty 
\left( \frac{\as}{\pi} \right)^n 
H_q^{F \,(n)}(x_1p_1, x_2p_2; {\bf\Omega})
\;\;, \\
\label{cqexp}
&C_{q \,a}(z;\as) = \delta_{q \,a} \;\,\delta(1-z) + 
\sum_{n=1}^\infty \left( \frac{\as}{\pi} \right)^n C_{q\, a}^{(n)}(z) \;\;.
\end{align}
The function $H_q^F$ is process dependent, whereas the functions
$C_{q \,a}$ are universal (they only depend on the parton indices). We add an
important remark on the factorized structure in the right-hand side
of Eq.~(\ref{what}): the scale of $\as$ is $M^2$ in the case of $H_q^F$, whereas
the scale is $b_0^2/b^2$ in the case of $C_{q \,a}$. The appearance of these two
different scales is essential \cite{Catani:2000vq} to disentangle the process
dependence of $H_q^F$ from the process-independent Sudakov form factor
($S_q$) and collinear functions ($C_{q \,a}$).

In the case of processes that are initiated at the Born level by the 
gluon fusion channel ($c=g$), the physics of the small-$q$ cross section
has a richer structure, which is the consequence of collinear correlations
\cite{Catani:2010pd}
that are produced by the evolution of the colliding hadrons into gluon partonic
states. In this case, the resummation formula (\ref{qtycross}) and, specifically,
its factor $\left[ H^F C_1 C_2 \right]$ are more involved than those
for the $q{\bar q}$ channel, since collinear radiation from the colliding gluons
leads to spin and azimuthal correlations 
\cite{Nadolsky:2007ba, Catani:2010pd}. The symbolic factor 
$\left[ H^F C_1 C_2 \right]$ in Eq.~(\ref{qtycross}) has the following
explicit form \cite{Catani:2010pd}:
\begin{align}
\label{whatgg}
\left[ H^F C_1 C_2 \right]_{gg;a_1a_2}
&= H_{g;\mu_1 \,\nu_1, \mu_2 \,\nu_2 }^F(x_1p_1, x_2p_2; {\bf\Omega}; \as(M^2))
\nn \\
&\times \; C_{g \,a_1}^{\mu_1 \,\nu_1}(z_1;p_1,p_2,{\bf b};\as(b_0^2/b^2)) 
\;\, C_{g \,a_2}^{\mu_2 \,\nu_2}(z_2;p_1,p_2,{\bf b};\as(b_0^2/b^2)) 
\;\;. 
\end{align}
where the function $H_{g}^{F}$ has the perturbative expansion
\begin{align}
\label{hexpgg}
H_{g}^{F\mu_1\nu_1,\mu_2\nu_2}(x_1p_1, x_2p_2; {\bf\Omega}; \as) 
=& H_{g}^{F(0)\mu_1\nu_1,\mu_2\nu_2}(x_1p_1, x_2p_2; {\bf\Omega})
\nn \\
+& \sum_{n=1}^\infty 
\left( \frac{\as}{\pi} \right)^n 
H_{g}^{F(n)\mu_1\nu_1,\mu_2\nu_2}(x_1p_1, x_2p_2; {\bf\Omega})
\;\;, 
\end{align}
and the following lowest-order normalization:
\begin{equation}
\label{h0gg}
H_{g}^{F(0)\mu_1\nu_1,\mu_2\nu_2} \;g_{\mu_1 \nu_1} 
\;g_{\mu_2 \nu_2} = 1 \;\;.
\end{equation}
Analogously to Eq.~(\ref{what}), in Eq.~(\ref{whatgg}) the function
$H_{g;\mu_1 \,\nu_1, \mu_2 \,\nu_2 }^F$ is process dependent (and it is
controlled by $\as$ at the scale $M^2$) and the partonic functions 
$C_{g a}^{\mu\,\nu}$ are process independent (and they are 
controlled by $\as$ at the scale $b_0^2/b^2$).
At variance with Eq.~(\ref{what}) (where the factorization structure in the
right-hand side is independent of the degrees of freedom of the colliding quark
and antiquarks), in Eq.~(\ref{whatgg})
the process-dependent function $H_{g}^F$ depends on the Lorentz indices
(and, thus, on the spins) $\{\mu_i \,\nu_i\}$ of the colliding gluons
with momenta $x_ip_i \;(i=1,2)$ and this dependence is coupled to (and correlated
with) a corresponding dependence of the partonic functions 
$C_{g a_i}^{\mu_i \,\nu_i}$.
The Lorentz tensor coefficients $C_{g a_i}^{\mu_i \,\nu_i}$ in Eq.~(\ref{whatgg})
depend on $b^2$ (through the scale of $\as$) and, moreover, they also
depend on the direction (i.e., the azimuthal angle) of the impact parameter
vector ${\bf b}$ in the transverse plane.
The structure of the partonic tensor is \cite{Catani:2010pd}
\begin{equation}
\label{cggten}
C_{g \,a}^{\,\mu \nu}(z;p_1,p_2,{\bf b};\as) =
d^{\,\mu \nu}(p_1,p_2) \;C_{g \,a}(z;\as) + D^{\,\mu \,\nu}(p_1,p_2;{\bf b}) 
\;G_{g \,a}(z;\as) \;\;,
\end{equation}
where
\begin{equation}
\label{dten}
d^{\,\mu \nu}(p_1,p_2) = - \,g^{\mu \nu} + 
\f{p_1^\mu p_2^\nu+ p_2^\mu p_1^\nu}{p_1 \cdot p_2} \;\;,
\end{equation}
\begin{equation}
\label{dbten}
D^{\,\mu \nu}(p_1,p_2;{\bf b}) = d^{\,\mu \nu}(p_1,p_2) - 
2 \; \f{b^\mu \,b^\nu}{\bf b^2} \;\;,
\end{equation}
and $b^\mu = (0,{\bf b},0)$ is the two-dimensional impact parameter vector
in the four-dimensional notation $(b^\mu b_\mu = - {\bf b^2})$.
The gluonic coefficient function $C_{g \,a}(z;\as)$ ($a=q,{\bar q},g$)
in the right-hand side of
Eq.~(\ref{cggten})
has the same perturbative structure as in Eq.~(\ref{cqexp}), and
it reads
\begin{equation}
\label{cgexp}
C_{g \,a}(z;\as) = \delta_{g \,a} \;\,\delta(1-z) + 
\sum_{n=1}^\infty \left( \frac{\as}{\pi} \right)^n C_{g\, a}^{(n)}(z) \;\;.
\end{equation}
In contrast,
the perturbative expansion of the coefficient functions $G_{ga}$, which are specific
to gluon-initiated processes, starts at ${\cal O}(\as)$, and we write
\begin{equation}
\label{gfexp}
G_{g \,a}(z;\as) = \f{\as}{\pi} \;G_{g \,a}^{(1)}(z) \,+
\sum_{n=2}^\infty \left( \frac{\as}{\pi} \right)^n G_{g \,a}^{(n)}(z) \;\;.
\end{equation}


We recall \cite{Catani:2010pd} an important physical consequence of the different
small-$q_T$ resummation structure between the $q{\bar q}$ annihilation 
and gluon fusion channels: the {\em absence} of {\em azimuthal correlations} 
with respect
to $\bqt$ in the $q{\bar q}$ annihilation channel, and the {\em presence} of
correlations with a definite predictable azimuthal dependence 
in the gluon fusion channel.
Indeed,
in the case of $q{\bar q}$ annihilation,
all the factors in the integrand of the Fourier
transformation on the right-hand side of the resummation formula
(\ref{qtycross}) are functions of ${\bf b}^2$, with no 
dependence on the azimuthal
angle $\phi({\bf b})$ 
of $\bf b$.
Therefore, the integration over $\phi({\bf b})$ in Eq.~(\ref{qtycross})
can be straightforwardly carried out,
and it leads \cite{Collins:1981uk, Collins:1984kg}
to a one-dimensional Bessel transformation that involves the 
$0$th-order Bessel function $J_0(b q_T)$.
This implies 
that the right-hand side of Eq.~(\ref{qtycross}) and, hence,
the singular part of the $\bqt$ differential cross 
section depend only on $q_T^2$,
with no additional dependence on the azimuthal angle $\phi(\bqt)$ 
of $\bqt$.
Unlike the case of $q{\bar q}$ annihilation,
the gluon fusion factor $\left[ H^F C_1 C_2 \right]$ in Eqs.~(\ref{qtycross})
and (\ref{whatgg}) does depend on the azimuthal
angle $\phi({\bf b})$ of the impact parameter $\bf b$.
Therefore, the integration over $\phi({\bf b})$ in the Fourier transformation
of Eq.~(\ref{qtycross}) is more complicated. It leads to one-dimensional Bessel
transformations that involve $J_0(b q_T)$ and higher-order Bessel functions,
such as the 2-nd order and 4th-order functions $J_2(b q_T)$ and $J_4(b q_T)$.
More importantly, it leads to a definite structure of azimuthal correlations
with respect to the azimuthal angle $\phi(\bqt)$ 
of the transverse momentum $\bqt$. The small-$q_T$ cross section in 
Eq.~(\ref{qtycross}) can be expressed \cite{Catani:2010pd} in terms of a
contribution that does not depend on $\phi(\bqt)$ plus a contribution that is
given by a linear combination of the four angular functions 
$\cos\left(2\phi(\bqt)\right), \sin\left(2\phi(\bqt)\right), 
\cos\left(4\phi(\bqt)\right)$ and $\sin\left(4\phi(\bqt)\right)$. 
No other functional dependence on $\phi(\bqt)$ is allowed by the resummation
formula (\ref{qtycross}) in the gluon fusion channel.

We recall that, 
due to its specific factorization structure,
the resummation formula in Eq.~(\ref{qtycross})
is invariant under the following renormalization-group
transformation \cite{Catani:2000vq}
\begin{align}
\label{Htrans}
H_c^F(\as)&\to H_c^F(\as) \left[h_c(\as)\right]^{-1}\, ,\\
\label{Btrans}
B_c(\as)&\to B_c(\as)-\beta(\as)\f{d\ln h_c(\as)}{d\ln \as}\, ,\\
C_{cb}(\as)&\to C_{cb}(\as)\left[h_c(\as)\right]^{1/2}\, ,
\label{Cscheme}
\end{align}
where $h_c(\as)=1+{\cal O}(\as)$ is an arbitrary perturbative function
(with $h_q(\as)=h_{\bar q}(\as)$).
More precisely, in the case of gluon-initiated processes, Eq.~(\ref{Cscheme})
becomes
\begin{equation}
\label{Cgscheme}
C^{\mu\nu}_{ga}(z;p_1,p_2,{\bf b};\as)\to C^{\mu\nu}_{ga}(\as)(z;p_1,p_2,{\bf
b};\as)\left[h_g(\as)\right]^{1/2}.
\end{equation}
In the right-hand side of Eq.~(\ref{Btrans}),
$\beta(\as)$ denotes the QCD $\beta$-function:
\begin{equation}
\f{d\ln\as(q^2)}{d\ln q^2}=\beta(\as(q^2))\, ,
\label{asevol}
\end{equation}
\begin{equation}
\beta(\as)=-\beta_0\,\as-\beta_1\,\as^2+{\cal O}(\as^3) \;,
\label{betaf}
\end{equation}
\begin{equation}
\label{betacoeff}
\beta_0=\f{11C_A-2N_f}{12\pi}~~,~~~~\beta_1=\f{17C_A^2-5C_AN_f-3C_FN_f}{24\pi^2}\; ,
\end{equation}
where $N_f$ is the number of quark flavours, $N_c$ is the number of colours, 
and the colour factors are $C_F=(N_c^2-1)/(2N_c)$ and $C_A=N_c$ in $SU(N_c)$ QCD.
As a consequence of the renormalization-group symmetry in 
Eqs.~(\ref{Htrans})--(\ref{Cgscheme}),
the resummation factors $H^F$, $S_c$, $C_{qa}$ and $C^{\mu\nu}_{ga}$ are
not {\em separately} defined (and, thus, computable) in an unambiguous way.
Equivalently, each of these separate factors can be precisely defined only by
specifying
a {\em resummation scheme} \cite{Catani:2000vq}.


To present the main results of this paper in the following sections, 
we find it convenient to specify a resummation scheme.
Therefore, in the rest of this paper we work in the scheme,
dubbed {\em hard scheme}, that is defined as follows.
The flavour off-diagonal coefficients $C^{(n)}_{ab}(z)$,
with $a\neq b$,
are `regular' functions of $z$ as $z\to 1$.
The $z$ dependence of 
the flavour diagonal coefficients $C^{(n)}_{qq}(z)$ and $C^{(n)}_{gg}(z)$
in Eqs.~(\ref{cqexp}) and (\ref{cgexp}) is instead due to both
`regular' functions and
`singular' distributions in the limit $z\to 1$. The 'singular' distributions are
$\delta(1-z)$ and the customary plus-distributions of the form 
$[(\ln^k(1-z))/(1-z)]_+\,$ ($k=0,1,2\dots$).
The {\em hard scheme} is the scheme in which,
order-by-order in perturbation theory, the coefficients $C^{(n)}_{ab}(z)$ with 
$n\geq 1$
do not contain any $\delta(1-z)$ term.
We remark (see also Sect.~\ref{sec:hardvirtual}) that this definition directly 
implies that all the process-dependent virtual corrections to the Born level
subprocesses in Eqs.~(\ref{qqpro}) and (\ref{ggpro})
are embodied in the resummation coefficient $H_c^F$.

We note that the specification of the hard scheme (or any other scheme)
has sole practical purposes of presentation
(theoretical results can be equivalently presented, as actually done in 
Refs.~\cite{Catani:2011kr} and \cite{Catani:2012qa}, by explicitly parametrizing
the resummation-scheme dependence of the resummation factors).
Having presented explicit results in the hard scheme, they can be translated
in other schemes by properly choosing the functions $h_c(\as)$ $(c=q,g)$ and
applying the transformation in Eqs.~(\ref{Htrans})--(\ref{Cgscheme}).
Moreover, and more importantly, the $q_T$ cross section, its all-order
resummation formula (\ref{qtycross}) and any consistent perturbative truncation
(either order-by-order in $\as$ or in classes of logarithmic terms) of the latter
\cite{Catani:2000vq, Bozzi:2005wk} are completely independent of the resummation
scheme.

The process-independent partonic coefficients $C_{ab}(z;\as)$ in Eqs.~(\ref{what})
and (\ref{whatgg}) are explicitly known up to the NNLO
(see references in Sect.~\ref{sec:procind}). The universality structure
of the process-dependent coefficients $H_c^F$ at NNLO and higher orders
(see Sect.~\ref{sec:hardvirtual})
is one of the main result of the present work.

\section{Process-independent coefficients}
\label{sec:procind}

Before discussing the general structure of the resummation coefficients
$H_c^F$,
in this section we present the expressions of the process-independent
resummation coefficients in the hard scheme, which is defined in Sec.~\ref{sec:resu}.

The partonic functions $C_{ab}$ and $G_{gb}$ in Eqs.~(\ref{cqexp}),
(\ref{cgexp}) and (\ref{gfexp}) depend on the parton indices. Owing to charge
conjugation invariance and flavour symmetry of QCD, the dependence on the parton
indeces is fully specified by the five independent quark functions
$\{ C_{qq}, C_{qq^\prime}, C_{q{\bar q}}, C_{q{\bar q}^\prime}, C_{qg} \}$
\cite{Catani:2012qa} ($q$ and $q^\prime$ denote quarks with different flavour)
and the four independent gluon functions
$\{ C_{gg}, C_{gq}, G_{gg}, G_{gq} \}$ \cite{Catani:2011kr}.

The first-order coefficients $C^{(1)}_{ab}(z)$ are explicitly known
\cite{Davies:1984hs,Kauffman:1991cx,deFlorian:2000pr,deFlorian:2001zd}.
Their expressions in the hard scheme can be obtained from
their corresponding  expression in an arbitrary scheme by simply setting the
coefficient of the $\delta(1-z)$ term to zero.
We get
\begin{align}
C^{(1)}_{qq}(z)&=\f{1}{2}C_F(1-z)\;,\\
C^{(1)}_{gq}(z)&=\f{1}{2}C_F\, z \;,\\
C^{(1)}_{qg}(z)&=\f{1}{2}z(1-z) \;,\\
C^{(1)}_{gg}(z)&= C_{q{\bar q}}(z) = C_{qq^\prime}(z) = C_{q{\bar q}^\prime}(z)
= 0\; .
\end{align}
The first-order coefficients $G_{ga}^{(1)}$ are resummation-scheme 
independent, and they
read \cite{Catani:2010pd}
\begin{equation}
G_{g \,a}^{(1)}(z) = C_a \;\f{1-z}{z}~~~~~~~~~~~a=q,g\, ,
\end{equation}
where $C_a$ is the Casimir colour coefficient of the parton $a$ with
$C_q=C_F$ and $C_g=C_A$.

According to Eq.~(\ref{Btrans}), the coefficients $B_a^{(n)}$ with $n \geq 2$ 
of the Sudakov form factor do depend \cite{Catani:2000vq}
on the resummation scheme.
The second-order process-independent
coefficient $B^{(2)}_c$ in Eq.~(\ref{abfun}) is known 
\cite{Davies:1984hs, deFlorian:2001zd}.
In the hard scheme, its value reads
\begin{equation}
B^{(2)}_a=\f{\gamma_{a(1)}}{16}+\pi\beta_0\, C_a\,\zeta_2\; ,
\end{equation}
where $\gamma_{a(1)}$ ($a=q,g$) are the coefficients
of the $\delta(1-z)$ term in the
NLO
quark and gluon splitting functions \cite{Curci:1980uw,Furmanski:1980cm},
which read
\begin{equation}
\gamma_{q\,(1)}
=\gamma_{{\bar q}\,(1)}
= (-3+24\zeta_2-48\zeta_3)\,C_F^2
+\left(-\frac{17}3-\frac{88}3\zeta_2 +24\zeta_3\right)\,C_F C_A
+\left(\frac{2}3+\frac{16}3\zeta_2 \right)\,C_F N_f\;,
\label{ga1q}
\end{equation}
\begin{equation}
\gamma_{g\,(1)}= \left(-\frac{64}3-24\zeta_3\right)\,C_A^2
+\frac{16}3\,C_A N_f
+4\,C_F N_f\;,
\label{ga1g}
\end{equation}
and $\zeta_n$ is the Riemann zeta-function ($\zeta_2=\pi^2/6,
\zeta_3=1.202\dots, \zeta_4=\pi^4/90$).

The second-order process-independent collinear coefficients $C_{ab}^{(2)}(z)$
of Eqs.~(\ref{cqexp}) and (\ref{cgexp})
have been computed in Refs.~\cite{Catani:2007vq, Catani:2009sm, Catani:2011kr,
Catani:2012qa}. The quark--quark coefficient $C_{qq}^{(2)}(z)$ has been 
independently computed in Ref.~\cite{Gehrmann:2012ze}.
The expressions of these coefficients 
in the hard scheme can be straightforwardly obtained from the results of Refs.~\cite{Catani:2011kr,Catani:2012qa} and are
explicitly reported below.

Starting from the quark channel, the coefficient $C^{(2)}_{qq}$ can be obtained
from Eq.~(34) of Ref.~\cite{Catani:2012qa}, and we have
\begin{equation}
2C^{(2)}_{qq}(z)={\cal H}^{DY(2)}_{q{\bar q}\ito q{\bar q}}(z)|_{{\rm
no}~\delta(1-z)}- \f{C_F^2}{4}\left[\left(2\pi^2-18\right)(1-z) - (1+z) \ln z\,
\right]\,,
\end{equation}
where ${\cal H}^{DY(2)}_{q{\bar q}\ito q{\bar q}}(z)|_{{\rm no}~\delta(1-z)}$
is obtained from the right-hand side of Eq.~(23) of Ref.~\cite{Catani:2012qa} by
setting the coefficient of the $\delta(1-z)$ term to zero.
Analogously, the coefficient $C^{(2)}_{qg}$ can be obtained from Eq.~(32) of
Ref.~\cite{Catani:2012qa} as
\begin{equation}
C^{(2)}_{qg}(z)={\cal H}^{DY(2)}_{q{\bar q}\ito qg}(z)-
\f{C_F}{4}\left[z\ln z+\f{1}{2}(1-z^2)+(\pi^2-8)z(1-z)\right]\, ,
\end{equation}
where ${\cal H}^{DY(2)}_{q{\bar q}\ito qg}(z)$
is given in Eq.~(27) of Ref.~\cite{Catani:2012qa}.
The flavour off-diagonal quark coefficients $C^{(2)}_{q{\bar q}}$, $C^{(2)}_{qq^\prime}$, $C^{(2)}_{q{\bar q}^\prime}$ are scheme independent and are presented in Eq.~(35) of Ref.~\cite{Catani:2012qa}.
Moving to the gluon channel, the coefficient $C_{gq}^{(2)}(z)$
can be obtained from Eq.~(32) of Ref.~\cite{Catani:2011kr}, and we have
\begin{equation}
2C_{gg}^{(2)}(z)={\cal H}^{H(2)}_{gg\ito gg}(z)|_{{\rm no}~\delta(1-z)}
+C_A^2 \left(\f{1+z}{z}\ln z+2\,\f{1-z}{z}\right)\, ,
\end{equation}
where ${\cal H}^{H(2)}_{gg\ito gg}(z)|_{{\rm no}~\delta(1-z)}$ is obtained
from the right-hand side Eq.~(24) of Ref.~\cite{Catani:2011kr} by setting the
coefficient of the $\delta(1-z)$ term to zero.
Finally, the coefficient $C_{gq}^{(2)}(z)$ can be obtained from
Eq.~(30) of Ref.~\cite{Catani:2011kr}, and we have
\begin{equation}
C_{gq}^{(2)}(z)={\cal H}^{H(2)}_{gg\ito gq}(z)
+C_F^2\f34 z+C_F C_A\f{1}{z}\left[(1+z)\ln z+2(1-z)-\f{5+\pi^2}{4}z^2\right]\;,
\end{equation}
where ${\cal H}^{H(2)}_{gg\ito gq}(z)$ is given in Eq.~(23) of Ref.~\cite{Catani:2011kr}.

The second-order gluon collinear coefficients $G_{ga}^{(2)}(z) \;(a=q,g)$
of Eq.~(\ref{gfexp}) are not yet known.
We can comment on the role of $G_{ga}^{(2)}$ in practical terms.
In the specific and important case of Higgs boson production by gluon fusion, the
coefficient $G_{ga}^{(2)}$ does not contribute to the cross section at the NNLO
(and NNLL accuracy). The Higgs boson cross section is discussed in detail in
Ref.~\cite{Catani:2010pd}: by direct inspection of Eq.~(45)
of Ref.~\cite{Catani:2010pd}, we can see that $G_{ga}^{(2)}$ starts to contribute
at the N$^3$LO. In most of the other processes (e.g., $F=\gamma \gamma, \;Z\gamma,
\;, W^+W^-$), the system $F$ can be produced by {\em both} $q{\bar q}$
annihilation and gluon fusion. In these case, due to the absence of direct
coupling of the gluons to the colourless particles in the system $F$, 
the production channel $gg \to F$ is suppressed by some powers of $\as$ with
respect to the channel $q{\bar q} \to F$. Therefore, also in these cases the
coefficient $G_{ga}^{(2)}$ does not contribute to the NNLO cross section.
This formal 
conclusion (based on counting the powers of $\as$) has a caveat, since the 
$gg \to F$ channel can receive a quantitative enhancement from the possibly large
luminosity of the gluon parton densities. However, the knowledge of the
first-order coefficients $C_{ga}^{(1)}$ and $G_{ga}^{(1)}$ should be sufficient to
compute the contribution from the channel $gg \to F$ to a quantitative level that
is comparable to that of the contribution from the channel $q{\bar q} \to F$
(whose collinear coefficients are fully known up to the second order).
In summary, we conclude that the effect of $G_{ga}^{(2)}$ rarely contributes in 
actual (practical)
computations of the $q_T$ cross section at the NNLO or NNLL accuracy.

\section{Hard-virtual coefficients}
\label{sec:hardvirtual}

In this section we focus on the process-dependent coefficient $H^F$.
In the hard scheme that we are using, this coefficient contains all the information on the
process-dependent virtual corrections, 
and, therefore,  
we can show that 
$H^F$ can be related in a process-independent (universal) way to the multiloop virtual amplitude
${\cal M}_{c{\bar c}\to F}$ of the partonic process $c{\bar c}\to F$.
In the following we first specify the notation that we use to denote the
all-loop virtual amplitude ${\cal M}_{c{\bar c}\to F}$. Then we introduce an
auxiliary (hard-virtual) amplitude $\widetilde{\cal M}_{c{\bar c}\to F}$
that is directly obtained from ${\cal M}_{c{\bar c}\to F}$ by using a
process-independent relation. Finally, we use the hard-virtual amplitude 
$\widetilde{\cal M}_{c{\bar c}\to F}$ to present the explicit expression of the 
hard-virtual coefficient $H^F$ up to the NNLO.

We consider the partonic {\em elastic}-production process
\begin{equation}
c({\hat p}_1)+ {\bar c}( {\hat p}_2)\to F(\{q_i\})\, ,
\label{partpro}
\end{equation}
where the two colliding partons with momenta ${\hat p}_1$ and ${\hat p}_2$
are either $c{\bar c}=gg$ or $c{\bar c}=q{\bar q}$ (we do not explicitly denote the flavour of the
quark $q$, although in the case with $c{\bar c}=q{\bar q}$, the quark and the
antiquark can have different flavours), and $F(\{q_i\})$ is the triggered 
final-state system in Eq.~(\ref{class}).
The loop scattering amplitude of the process in Eq.~(\ref{partpro})
contains ultraviolet (UV) and infrared (IR) singularities,
which are regularized in $d=4-2\epsilon$ space-time dimensions by using the customary scheme of
conventional dimensional regularization (CDR)\footnote{The relation between the
CDR scheme and other variants of dimensional regularization is explicitly known
\cite{dimreg} up to the two-loop level.}.
Before performing renormalization, the multiloop QCD amplitude 
has a perturbative dependence on
powers of $\as^u\mu_0^{2\ep}$, where $\as^u$ is the bare coupling and $\mu_0$ is
the dimensional-regularization scale. In the following we work with the renormalized 
on-shell scattering amplitude
that
is obtained from the corresponding unrenormalized amplitude by just expressing the bare coupling $\as^u$
in terms of the running coupling $\as(\mu_R^2)$ according to the \ms\ scheme
relation
\begin{equation}
\as^u \,\mu_0^{2\ep}S_\ep=\as(\mu_R^2) \,\mu_R^{2\ep}\left[1-\as(\mu_R^2)\f{\beta_0}{\ep}+\as^2(\mu_R^2)\left(\f{\beta_0^2}{\ep^2}-\f{\beta_1}{2\ep}\right)+{\cal O}(\as^3(\mu_R^2))\right]
\;\;,
\end{equation}
where $\mu_R$ is the renormalization scale, $\beta_0$ and $\beta_1$ are the first two coefficients of the QCD $\beta$-function in Eq.~(\ref{betacoeff})
and the factor $S_\ep$ is
\begin{equation}
S_\ep=(4\pi)^\ep \,e^{-\ep\gamma_E}\, .
\end{equation}
The renormalized all-loop amplitude of the 
process in Eq.~(\ref{partpro}) is
denoted by ${\cal M}_{c{\bar c}\to F}({\hat p}_1, {\hat p}_2;\qi)$,
and it has the perturbative (loop) expansion
\begin{align}
{\cal M}_{c{\bar c}\to F}({\hat p}_1, {\hat p}_2;\qi) \!
&= \left( \as(\mu_R^2) \,\mu_R^{2\ep} \right)^k
\left[
{\cal M}_{c{\bar c}\to F}^{\,(0)}({\hat p}_1, {\hat p}_2; \qi)
+\left( \frac{\as(\mu_R^2)}{2\pi}\right) 
{\cal M}_{c{\bar c}\to F}^{\,(1)}({\hat p}_1, {\hat p}_2; \qi; \mu_R)
\right.\nn\\
&+\left.\!\!\left(\frac{\as(\mu_R^2)}{2\pi}\right)^{\!\!2}
\!\!{\cal M}_{c{\bar c}\to F}^{\,(2)}({\hat p}_1, {\hat p}_2; \qi; \mu_R)
\!+ \sum_{n=3}^{\infty} \left(\frac{\as(\mu_R^2)}{2\pi}\right)^{\!\!n}
\!\!{\cal M}_{c{\bar c}\to F}^{\,(n)}({\hat p}_1, {\hat p}_2; \qi; \mu_R)
\right] ,
\label{ampli}
\end{align}
where the value $k$ of the overall power of $\as$ depends on the specific
process
(for instance, $k=0$ in the case of the vector boson production 
process $q{\bar q}\to V$, and $k=1$ 
in the case of the Higgs boson production process $gg \to H$
through a heavy-quark loop).
Note also that the lowest-order perturbative term 
${\cal M}_{c{\bar c}\to F}^{\,(0)}$ 
is not necessarily a tree-level amplitude
(for instance, it involves a quark loop in the cases $gg \to H$ and $gg \to
\gamma \gamma$). The perturbative terms ${\cal M}_{c{\bar c}\to F}^{\,(l)}$
$(l=1,2,\dots)$
are UV finite, but they still depend on $\ep$ 
(although this dependence is not explicitly denoted in Eq.~(\ref{ampli})). 
In particular, the amplitude ${\cal M}_{c{\bar c}\to F}^{\,(l)}$
at the $l$-th perturbative order is IR divergent as $\ep\to 0$, and it behaves as
\begin{equation}
{\cal M}_{c{\bar c}\to F}^{\,(l)}\sim \left(\f{1}{\ep}\right)^{2l}+\dots\, ,
\end{equation}
where the dots stand for $\ep$-poles of lower order.
The IR divergent contributions to the scattering amplitude have a universal
structure \cite{Catani:1998bh}, which is explicitly known at the 
one-loop \cite{ir1loop, Catani:1998bh},
two-loop \cite{Catani:1998bh, ggFF2} 
and three-loop \cite{FF3, 3loopsing} level for the class of processes in
Eq.~(\ref{partpro}).

The explicit calculations and the results of Ref.~\cite{deFlorian:2001zd} 
show that the
NLO hard-virtual coefficient $H^{F\,(1)}$ is explicitly related in a
process-independent form to the leading-order (LO) amplitude ${\cal M}_{c{\bar c}\to F}^{\,(0)}$
and to the IR finite part of the NLO amplitude 
${\cal M}_{c{\bar c}\to F}^{\,(1)}$.
The relation between $H^F_c$ and ${\cal M}_{c{\bar c}\to F}$ can be extended to
the NNLO and to higher-order levels. This extension can be formulated and
expressed in simple and general terms by introducing an 
auxiliary (hard-virtual) amplitude $\widetilde{\cal M}_{c{\bar c}\to F}$
that is directly obtained from ${\cal M}_{c{\bar c}\to F}$ in a universal
(process-independent) way. In practice, $\widetilde{\cal M}_{c{\bar c}\to F}$
is obtained from ${\cal M}_{c{\bar c}\to F}$ by removing its IR divergences and a
{\em definite} amount of IR finite terms. The (IR divergent and finite) terms
that are removed from ${\cal M}_{c{\bar c}\to F}$ originate from real emission
contributions to the cross section and, therefore, these terms and 
$\widetilde{\cal M}_{c{\bar c}\to F}$ {\em specifically}
depend on the
transverse-momentum cross section of Eq.~(\ref{diffxs}), which we consider 
throughout this paper.

The hard-virtual amplitude $\widetilde{\cal M}_{c{\bar c}\to F}$ has a
perturbative expansion that is analogous to that in Eq.~(\ref{ampli}). We 
write
\begin{align}
\widetilde{\cal M}_{c{\bar c}\to F}({\hat p}_1, {\hat p}_2;\qi) \!
&= \left( \as(\mu_R^2) \,\mu_R^{2\ep} \right)^k
\left[
\widetilde{\cal M}_{c{\bar c}\to F}^{\,(0)}({\hat p}_1, {\hat p}_2; \qi)
+\left( \frac{\as(\mu_R^2)}{2\pi}\right) 
\widetilde{\cal M}_{c{\bar c}\to F}^{\,(1)}({\hat p}_1, {\hat p}_2; \qi; \mu_R)
\right.\nn\\
&+\left.\!\!\left(\frac{\as(\mu_R^2)}{2\pi}\right)^{\!\!2}
\!\!\widetilde{\cal M}_{c{\bar c}\to F}^{\,(2)}({\hat p}_1, {\hat p}_2; \qi; \mu_R)
\!+ \sum_{n=3}^{\infty} \left(\frac{\as(\mu_R^2)}{2\pi}\right)^{\!\!n}
\!\!\widetilde{\cal M}_{c{\bar c}\to F}^{\,(n)}({\hat p}_1, {\hat p}_2; \qi; \mu_R)
\right] .
\label{subampli}
\end{align}
At the LO, 
$\widetilde{\cal M}_{c{\bar c}\to F}$ and ${\cal M}_{c{\bar c}\to F}$ coincide,
and we have
\begin{equation}
\widetilde{\cal M}_{c{\bar c}\to F}^{(0)}={\cal M}_{c{\bar c}\to F}^{(0)}\, .
\label{ampli0}
\end{equation}
At higher-perturbative orders, $\widetilde{\cal M}_{c{\bar c}\to F}^{\,(l)}$ is
expressed in terms of the amplitudes ${\cal M}_{c{\bar c}\to F}^{\,(n)}$
at equal or lower orders (i.e. with $n \leq l$). At NLO and NNLO
(see Sect.~\ref{hvstruct} for higher-order terms), we explicitly
have
\begin{equation}
\widetilde{\cal M}_{c{\bar c}\to F}^{(1)}={\cal M}_{c{\bar c}\to F}^{(1)}-
\tilde{I}_c^{(1)}(\ep,M^2/\mu_R^2) \;\, {\cal M}_{c{\bar c}\to F}^{(0)}\; ,
\label{ampli1}
\end{equation}
\begin{equation}
\widetilde{\cal M}_{c{\bar c}\to F}^{(2)}=\mathcal{M}_{c{\bar c}\to F}^{(2)}
-\tilde{I}_c^{(1)}(\ep,M^2/\mu_R^2) \;
\mathcal{M}_{c{\bar c}\to F}^{(1)}-
\tilde{I}^{(2)}_c(\ep,M^2/\mu_R^2) \;\mathcal{M}_{c{\bar c}\to F}^{(0)}\, .
\label{ampli2}
\end{equation}
In Eqs.~(\ref{ampli1}) and (\ref{ampli2}) the functional dependence of
the perturbative amplitudes on their argument 
$({\hat p}_1, {\hat p}_2; \qi; \mu_R)$ is not explicitly 
recalled. 
The perturbative terms $\tilde{I}_c^{(1)}$ and $\tilde{I}_c^{(2)}$ act as
IR subtraction operators (factors), and their functional dependence is explicitly
denoted in Eqs.~(\ref{ampli1}) and (\ref{ampli2}). These terms are process
independent (they do not depend on $F$ and on its specific production mechanism
in Eq.~(\ref{partpro})): they only depend on the invariant mass $M$ of the
system $F$ (through the dimensionless ratio $M^2/\mu_R^2$), on the type $c$
$(c=q,g)$ of colliding partons, and on $\ep$. In particular, 
$\tilde{I}_c^{(1)}$ and $\tilde{I}_c^{(2)}$ include $\ep$-pole contributions
that cancel the IR divergences of ${\cal M}_{c{\bar c}\to F}^{(1)}$ and
${\cal M}_{c{\bar c}\to F}^{(2)}$, so that the hard-virtual amplitudes
$\widetilde{\cal M}_{c{\bar c}\to F}^{(1)}$ and
$\widetilde{\cal M}_{c{\bar c}\to F}^{(2)}$ are IR finite as $\ep \to 0$.
We also note that the structure of Eqs.~(\ref{ampli1}) and (\ref{ampli2})
and the explicit dependence of $\tilde{I}_c^{(1)}$ and $\tilde{I}_c^{(2)}$
on $M^2/\mu_R^2$ guarantee (see Eqs.~(\ref{i1til}), (\ref{i2til}) and 
Sect.~\ref{hvstruct}) that the hard-virtual amplitude 
$\widetilde{\cal M}_{c{\bar c}\to F}$ is renormalization-group invariant
(analogously to ${\cal M}_{c{\bar c}\to F}$).

The explicit expression of the first-order (one-loop) subtraction operator
$\tilde{I}_a^{(1)}$ is
\begin{equation}
\tilde{I}_a^{(1)}(\epsilon,M^2/\mu_R^2)
= \tilde{I}^{(1)\,{\rm soft}}_{a}(\epsilon,M^2/\mu_R^2)
+\tilde{I}^{(1)\,{\rm coll}}_{a}(\epsilon,M^2/\mu_R^2)\;,
\label{i1til}
\end{equation}
with
\begin{align}
\tilde{I}^{(1)\,{\rm soft}}_{a}(\epsilon,M^2/\mu_R^2)
&= -\frac{e^{\epsilon\gamma_E}}{\Gamma(1-\epsilon)}\,
\left(\frac{1}{\epsilon^2}
+ i \pi \,\frac{1}{\epsilon}
+\delta^{q_T}\right)\,
C_a \left( \frac{M^2}{\mu_R^2}\right)^{-\epsilon}\;,
\label{i1soft} \\
\tilde{I}^{(1)\,{\rm coll}}_{a}(\epsilon,M^2/\mu_R^2)
&=-\frac1\epsilon\;\gamma_a\left(\frac{M^2}{\mu_R^2}\right)^{-\epsilon}\;,
\label{i1coll}
\end{align}
and
\begin{equation}
\gamma_q=\gamma_{\bar q}=\frac32 \,C_F\,, \qquad
\gamma_g=\frac{11}6 \,C_A -\frac13 \,N_f\;.
\label{galowest}
\end{equation}
The coefficient $\delta^{q_T}$ affects only the IR finite part of the subtraction operator. The known results on the NLO hard-collinear coefficients 
$H_c^{F\,(1)}$
\cite{deFlorian:2001zd}
are recovered by fixing
\begin{equation}
\delta^{q_T}=0\, .
\label{delowest}
\end{equation}

The second-order (two-loop) subtraction operator $\tilde{I}^{(2)}_c$ is
\begin{align}
\tilde{I}_a^{(2)}(\epsilon,M^2/\mu_R^2)
=&-\frac12
\left[\tilde{I}_a^{(1)}(\epsilon,M^2/\mu_R^2)
\right]^2\!
+\left\{
\frac{2\pi\beta_0}{\epsilon}
\bigg[
\tilde{I}_a^{(1)}(2\epsilon,M^2/\mu_R^2)
\right. \nn \\ 
&-\left.
\tilde{I}_a^{(1)}(\epsilon,M^2/\mu_R^2)
\bigg]
+
K\, \tilde{I}_a^{(1)\,{\rm soft}}(2\epsilon,
M^2/\mu_R^2)
+
\widetilde{H}_a^{(2)}(\epsilon,M^2/\mu_R^2)
\right\}\;,
\label{i2til}
\end{align}
with
\begin{align}
\widetilde{H}_a^{(2)}(\epsilon,M^2/\mu_R^2)
&= \widetilde{H}_a^{(2)\,{\rm coll}}(\epsilon,M^2/\mu_R^2) 
+\widetilde{H}_a^{(2)\,{\rm soft}}(\epsilon,M^2/\mu_R^2) 
\label{h2sc}\\
&= \frac1{4\epsilon}\left(\frac{M^2}{\mu_R^2}\right)^{-2\epsilon}
\left(\frac14 \,\gamma_{a\,(1)}  
+\,C_a \,d_{(1)}
+\epsilon \,C_a \,\delta_{(1)}^{q_T}
\right)\;,
\label{h2expl} 
\end{align}
where $\widetilde{H}_a^{(2)\,{\rm coll}}$ is the contribution that is proportional
to $\gamma_{a\,(1)}$ and $\widetilde{H}_a^{(2)\,{\rm soft}}$ is the remaining
contribution (which is proportional to $C_a$) in Eq.~(\ref{h2expl}).
The QCD coefficients $K$ in Eq.~(\ref{i2til}) 
and $d_{(1)}$ in Eq.~(\ref{h2expl}) (they control the IR divergences of
$\tilde{I}_a^{(2)}$) are \cite{Catani:1998bh}
\begin{equation}
K=\left(\frac{67}{18}-\frac{\pi^2}{6}\right)\,C_A-\frac{5}9\,N_f \;,
\end{equation}
\begin{equation}
d_{(1)} = \left(\frac{28}{27} - \frac{1}{3} \zeta_2 \right) N_f
+ \left( -\frac{202}{27} + \frac{11}{6} \zeta_2 + 7 \zeta_3
\right) C_A \,,
\label{donecoef}
\end{equation}
and the coefficients $\gamma_{a\,(1)}$ ($a=q,g$) are given in Eqs.~(\ref{ga1q})
and (\ref{ga1g}).
The coefficient $\delta^{q_T}_{(1)}$ in Eq.~(\ref{h2expl})
affects only the IR finite part of the two-loop subtraction operator. We find
(see Sect.~\ref{hvstruct})
\begin{equation}
\delta^{q_T}_{(1)}=\f{20}{3} \zeta_3\pi
\beta_0+\left(-\f{1214}{81}+ \f{67}{18} \zeta_2\right) C_A+ \left( \f{164}{81} -
\f{5}{9} \zeta_2
\right) N_f\, .
\label{delnnlo}
\end{equation}

Having introduced the subtracted amplitude $\widetilde{\cal M}_{c{\bar c}\to F}$,
we can relate it to the process-dependent resummation coefficients $H^F_c$ 
of Eqs.~(\ref{qtycross}), (\ref{what}) and (\ref{whatgg}).
In the case of processes initiated by $q{\bar q}$ annihilation
(see Eqs.~(\ref{what}) and (\ref{hexp})), 
the {\em all-order} coefficient $H^F_q$ can be written as
\begin{equation}
\label{Hq}
\as^{2k}(M^2) \,H^F_q(x_1p_1, x_2p_2; {\bf\Omega};\as(M^2))
=\f{|\widetilde{\cal M}_{q{\bar q}\to F}(x_1p_1, x_2p_2;\qi)|^2}{|{\cal M}_{q{\bar
q}\to F}^{(0)}(x_1p_1, x_2p_2;\qi)|^2}\, ,
\end{equation}
where $k$ is the value of the overall power of $\as$ in the expansion of 
${\cal M}_{c{\bar c}\to F}$ (see Eqs.~(\ref{ampli}) and (\ref{subampli})).
In the case of processes initiated by gluon fusion 
(see Eqs.~(\ref{whatgg}) and (\ref{hexpgg})),
the analogue of Eq.~(\ref{Hq}) is 
\begin{equation}
\label{Hg}
\as^{2k}(M^2) 
\,h_g^{F\,\mu_1\nu_1\mu_2\nu_2}(x_1p_1, x_2p_2; {\bf\Omega};\as(M^2))
=\f{\left[\widetilde{\cal M}^{\mu_1\mu_2}_{gg\to F}(x_1p_1,
x_2p_2;\qi)\!\right]^\dagger\!\widetilde{\cal M}^{\nu_1\nu_2}_{gg\to F}(x_1p_1,
x_2p_2;\qi)}{|{\cal M}^{(0)}_{gg\to F}(x_1p_1, x_2p_2;\qi)|^2}\,,
\end{equation}
and the {\em all-order} coefficient $H^F_g$ is \cite{Catani:2010pd}
\begin{equation}
\label{Hgtens}
H_g^{F\mu_1\nu_1\mu_2\nu_2}(x_1p_1, x_2p_2; {\bf\Omega};\as)=
d^{\,\mu_1}_{\,\;\mu_1^\prime} \;d^{\,\nu_1}_{\,\;\nu_1^\prime} \;
d^{\,\mu_2}_{\,\;\mu_2^\prime}\;d^{\,\nu_2}_{\,\;\nu_2^\prime} \;
h_g^{F\,\mu_1^\prime\nu_1^\prime\mu_2^\prime\nu_2^\prime}(x_1p_1, x_2p_2;
{\bf\Omega};\as) \,,
\end{equation}
where $d^{\,\mu \nu}= d^{\,\mu \nu}(p_1,p_2)$ is the polarization tensor in
Eq.~(\ref{dten}) 
and it projects onto the Lorentz indices in the transverse plane.

In Eqs.~(\ref{Hq}) and (\ref{Hg}), the notation 
$|\widetilde{\cal M}_{q{\bar q}\to F}|^2$ and $|{\cal M}^{(0)}_{gg\to F}|^2$
denotes the squared amplitudes {\em summed} over the colours of the colliding
partons and over the ({\em physical}) spin polarization states of the colliding partons and of
the particles in the final-state system $F$. In the numerator of
Eq.~(\ref{Hg}), the sum over the spin polarization states of the initial-state 
gluons is not performed, and the amplitude  
$\widetilde{\cal M}^{\nu_1\nu_2}_{gg\to F}$ depends on the Lorentz index $\nu_i$
$(i=1,2)$ of the incoming gluon leg with momentum $x_ip_i$. 
The Lorentz indices of $\widetilde{\cal M}_{gg\to F}$ coincide with those of
${\cal M}_{gg\to F}$ in Eqs.~(\ref{ampli0})--(\ref{ampli2}),
since the subtraction factors $\tilde{I}_c^{(1)}$ and $\tilde{I}_c^{(2)}$
do not depend on the spin.
We recall that, according to the notation in Eq.~(\ref{diffxs}), the kinematics
of the final-state momenta $\{q_i\}$ is fully specified by the total momentum
$q=\sum_i q_i$ and the set of variables ${\bf\Omega}$. Therefore, the dependence
of the amplitudes on $\{q_i\}$ completely determines the ${\bf\Omega}$
dependence of $H^F_c$ in Eqs.~(\ref{Hq}) and (\ref{Hg}).
To be precise, we also note that $H^F_c$ is computed in $d=4$ space-time
dimensions and, therefore, the right-hand side of 
Eqs.~(\ref{Hq}) and (\ref{Hg}) has to be evaluated in the limit $\ep \to 0$
(this limit is well-defined and straightforward, since 
${\cal M}^{(0)}_{c{\bar c}\to F}$ and the order-by-order expansion of the
hard-virtual amplitude $\widetilde{\cal M}_{c{\bar c}\to F}$ are IR finite).
An additional remark regards the dependence on the renormalization scale.
According to the resummation formula (\ref{qtycross}), the {\em all-order}
factor 
$d\sigma_{c{\bar c}, \,F}^{(0)} \,H^F_c$
(and, consequently, the left-hand side of Eqs.~(\ref{Hq}) and (\ref{Hg}))
is renormalization-group invariant, and it is perturbatively computable as
series expansion in powers of $\as(M^2)$, with no dependence on $\mu_R$.
This property is fully consistent with the form of
Eqs.~(\ref{Hq}) and (\ref{Hg}), since the all-order hard-virtual amplitude in
the right-hand side of these equations is a renormalization-group invariant
quantity. Obviously, each side of these equations can be expanded in powers of
$\as(\mu_R^2)$, thus leading to corresponding perturbative coefficients that
explicitly depend on $M^2/\mu_R^2$.

The expressions (\ref{Hq}), (\ref{Hg}), (\ref{ampli1}) and (\ref{ampli2})
and the explicit results in Eqs.~(\ref{i1til}) and (\ref{i2til}) permit the
straightforward computation of the process-dependent resummation coefficients 
$H^F_c$ for an arbitrary process of the class in Eq.~(\ref{class}). The
explicit computation of $H^F_c$ up to the NNLO is elementary, 
provided the scattering amplitude ${\cal M}_{c{\bar c}\to F}$
of the corresponding partonic subprocess is available (known) up to the
NNLO (two-loop) level. Some examples are explicitly reported in 
Appendix~\ref{appa}.
In particular, in Appendix~\ref{appa} we use Eqs.~(\ref{ampli0})--(\ref{ampli2})
and (\ref{Hq}), and we present the explicit expression of the NNLO hard-virtual
coefficient $H^{\gamma \gamma (2)}_q$ for the process of diphoton production
\cite{Catani:2011qz}.

\subsection{The structure of the hard-virtual term}
\label{hvstruct}

The all-loop amplitude ${\cal M}_{c{\bar c}\to F}$ receives contributions from
loop momenta in different kinematical regions. Roughly speaking, the loop
momentum $k$ can be in the UV region ($k \gg M$), in the IR region ($k \ll M$)
or in the hard (intermediate) region ($k\sim M$). The UV contributions are
treated and `removed' by renormalization. As already anticipated (and discussed
below), the subtraction factor $\tilde{I}_c$ has a IR (soft and collinear)
origin, and it `removes' the IR contributions from ${\cal M}_{c{\bar c}\to F}$.
The subtracted (hard-virtual) amplitude 
$\widetilde{\cal M}_{c{\bar c}\to F}$ can thus be interpreted as originating from
the hard component of the virtual radiative corrections to the LO amplitude 
${\cal M}_{c{\bar c}\to F}^{(0)}.$

The iterative structure of Eqs.~(\ref{ampli0})--(\ref{ampli2}) can be recast in
factorized form. We have
\begin{equation}
\widetilde{\cal M}_{c{\bar c}\to F}({\hat p}_1, {\hat p}_2;\qi) =
\left[ 1 - \tilde{I}_c(\ep,M^2) \right]  
{\cal M}_{c{\bar c}\to F}({\hat p}_1, {\hat p}_2;\qi) \;\;,
\label{hvall}
\end{equation}
with
\begin{equation}
\tilde{I}_c(\ep,M^2) = \frac{\as(\mu_R^2)}{2\pi}
\,\tilde{I}_c^{(1)}(\ep,M^2/\mu_R^2) + 
\left( \frac{\as(\mu_R^2)}{2\pi} \right)^2 \tilde{I}_c^{(2)}(\ep,M^2/\mu_R^2)
+ \sum_{n=3}^\infty \left( \frac{\as(\mu_R^2)}{2\pi} \right)^n
\tilde{I}_c^{(n)}(\ep,M^2/\mu_R^2) \,.
\label{itilall}
\end{equation}
The {\em factorization} formula (\ref{hvall}) gives the {\em all-order} relation
between the hard-virtual amplitude $\widetilde{\cal M}_{c{\bar c}\to F}$
(which determines $H_c^F$ through Eqs.~(\ref{Hq}) and (\ref{Hg}))
and the scattering amplitude ${\cal M}_{c{\bar c}\to F}$ (the perturbative
expansion of Eqs.~(\ref{hvall}) up to the NNLO exactly gives 
Eqs.~(\ref{ampli0})--(\ref{ampli2})). The all-order subtraction factor 
$\tilde{I}_c(\ep,M^2)$ in Eq.~(\ref{hvall}) is independent of $\mu_R$ and,
thus, 
it is renormalization-group invariant. The order-by-order dependence
on $\mu_R$ simply arises from the expansion (see Eq.~(\ref{itilall}))
in terms of powers of $\as(\mu_R^2)$ and perturbative coefficients 
$\tilde{I}_c^{(n)}(\ep,M^2/\mu_R^2)$. Note that $\tilde{I}_c(\ep,M^2)$
depends on $\ep$, and we are referring to renormalization-group invariance in
$d=4-2\ep$ dimensions (i.e., to all orders in $\ep$), where the right-hand side
of Eq.~(\ref{asevol})
has to be modified by replacing $\beta(\as)$ with the $d$-dimensional
$\beta$-function $\beta(\ep,\as) = - \,\ep +\beta(\as)$. The $\mu_R$ independence
of $\tilde{I}_c(\ep,M^2)$ can be explicitly checked up to the NNLO by using the
expressions of $\tilde{I}_c^{(1)}(\ep,M^2/\mu_R^2)$ and
$\tilde{I}_c^{(2)}(\ep,M^2/\mu_R^2)$ in Eqs.~(\ref{i1til})--(\ref{h2expl}).

We can illustrate the origin and the derivation of the results in 
Eqs.~(\ref{Hq}), (\ref{Hg}) and (\ref{hvall}) by starting from the direct
computation of the $q_T$ cross section in Eq.~(\ref{diffxs}).
The calculation of the cross section or, more precisely, of the corresponding
partonic cross sections involves three types of contributions: 
(i) the elastic-production process in Eq.~(\ref{partpro}); 
(ii) inelastic (real-emission) processes, where the
system $F$ is accompanied by additional final-state partons;
(iii) the collinear counterterm that is necessary to define the \ms\
parton densities in terms of the bare (na\"ive) parton densities.

(i) The elastic process directly contributes to the partonic cross section and
thus to $H_c^F$ with a term that is proportional to (the square of) the all-loop
amplitude ${\cal M}_{c{\bar c}\to F}$.

(ii) Since we are interested in the small-$q_T$ singular cross section
of Eq.~(\ref{qtycross}),
the calculation of the inelastic processes can be simplified. 
In the small-$q_T$ limit, the additional final-state partons in the inelastic 
processes must be either soft or collinear to one of the colliding partons
(non-soft and non-collinear partons give cross section contributions that are
relatively suppressed by some powers
of $q_T/M \sim 1/(bM)$).
The radiation of soft \cite{Catani:1999ss, Bern:1999ry, Catani:2000pi}
and collinear \cite{Campbell:1997hg, Catani:1999ss, Bern:1999ry, Kosower:1999rx} 
partons from two colliding partons is described by QCD factorization formulae,
where the singular soft/collinear term (which includes its virtual radiative
corrections) is universal (process independent) and it acts onto the all-loop
amplitude ${\cal M}_{c{\bar c}\to F}$ as in the factorized expression on the
right-hand side of Eq.~(\ref{hvall}).
To be precise, soft/collinear factorization works at the amplitude (and squared
amplitude) level, and it can be spoiled by kinematical effects at the cross
section level, i.e., after the (cross section dependent) phase-space integration
of the squared amplitudes. However, in our case, factorization breaking effects of
kinematical origin cannot arise, since we are effectively working in impact
parameter space (in the small-$q_T$ limit, the kinematics of the 
$q_T$ cross section is exactly factorized \cite{Parisi:1979se} by the 
Fourier transformation
to $\bf b$ space). More precisely, we are considering the cross section
contributions at fixed values of the impact
parameter $\bf b$, namely, the integrand terms on the right-hand side of
Eq.~(\ref{qtycross}): the factorized structure of these terms directly follows
form soft/collinear factorization formulae.
The following step in our discussion consists in the observation that the
radiation of collinear or, more precisely, non-soft collinear partons requires a
non-vanishing longitudinal-momentum recoil and, therefore, it cannot contribute
to the factor $H_c^F$ in Eq.~(\ref{qtycross}) (non-soft collinear radiation
definitely contribute to the other $\bf b$-dependent factors on the right-hand
side of Eq.~(\ref{qtycross})). In summary, considering the inelastic processes,
the factor $H_c^F$ receives contributions only from {\em soft} radiation:
these are the factorized soft contributions that are $\bf b$-independent
and that do not vanish in the near-elastic limit.

(iii) The calculation of the elastic and inelastic processes gives the bare
partonic cross section. The introduction of the (IR divergent)
{\em collinear} counterterm of the parton densities amounts to multiply the entire
bare partonic cross section with a process-independent factor that has a
convolution structure with respect to the longitudinal-momentum fractions of the
colliding partons. The evolution kernel of the convolution is the
Altarelli--Parisi splitting function $P_{ab}(z;\as)$, and only 
the soft part (analogously to the contribution of the inelastic processes) and
the {\em virtual}
part (i.e., the part that is proportional to $\delta(1-z)$) of the splitting
function can contribute to the factor $H_c^F$ in Eq.~(\ref{qtycross}).
This `virtual-collinear' contribution to $H_c^F$ is process independent
(it only depends on the type, i.e. quarks or gluons, of colliding partons), and
it has the same factorized structure as in Eq.~(\ref{hvall}). Note that the
explicit form of the collinear counterterm is fully specified
\cite{Curci:1980uw}, since we are
considering parton densities defined in the \ms\ factorization scheme. 
Moreover, we can add that the {\em entire} virtual part of the collinear
counterterm is included in $H_c^F$, since we are working in the {\em hard scheme}
(where the perturbative corrections to the coefficients $C_{ab}(z;\as)$
of Eq.~(\ref{qtycross}) contain no contributions that are proportional to 
$\delta(1-z)$).

In summary, from our general discussion we can conclude that, in the hard scheme,
the resummation factor $H_c^F$ has the structure given by 
Eqs.~(\ref{Hq}), (\ref{Hg}) and (\ref{hvall}). The all-loop amplitude 
${\cal M}_{c{\bar c}\to F}$ of Eq.~(\ref{hvall}) originates from the elastic
process in Eq.~(\ref{partpro}). The remaining universal factor 
$\tilde{I}_c(\ep,M^2)$ on the right-hand side of Eq.~(\ref{hvall}) includes two
types of contributions: a soft contribution (from inelastic processes and the
collinear counterterm) and a
collinear contribution from the virtual part of the \ms\ collinear counterterm.
In the following, we combine these conclusions with two additional properties
of  $H_c^F$ (its renormalization-group invariance and its IR finiteness), and we
shall show that the explicit form of $\tilde{I}_c(\ep,M^2)$ is (almost)
completely determined up to the NNLO.

The IR divergences of the all-loop amplitude ${\cal M}_{c{\bar c}\to F}$
have a known universal structure that can be presented in the following form
\cite{Catani:1998bh}:
\begin{equation}
{\cal M}_{c{\bar c}\to F} = I_c \;{\cal M}_{c{\bar c}\to F}
+   {\cal M}_{c{\bar c}\to F}^{\rm fin.} \;\;,
\label{irdiv}
\end{equation}
where the all-loop factor $I_c$ has a perturbative expansion that is analogous to
that in Eq.~(\ref{itilall}). The component 
${\cal M}_{c{\bar c}\to F}^{\rm fin.}$ of the amplitude is IR finite as $\ep \to
0$, while the process-independent factor $I_c$ includes IR divergent $\ep$-poles.
The perturbative (loop) expansion of Eq.~(\ref{irdiv}) iteratively determines the
IR divergent component of ${\cal M}_{c{\bar c}\to F}$ (see Eqs.~(12) and (18) in
Ref.~\cite{Catani:1998bh}). Note that the separation between 
${\cal M}_{c{\bar c}\to F}^{\rm fin.}$ and IR divergent terms depends on the
amount of IR finite contributions that are actually included in 
$I_c$ (in particular, as discussed below, the hard-virtual amplitude
$\widetilde{\cal M}_{c{\bar c}\to F}$ can be regarded as a specific definition of
${\cal M}_{c{\bar c}\to F}^{\rm fin.}$).

The relation (\ref{irdiv}) can be rewritten as
\begin{equation}
{\cal M}_{c{\bar c}\to F}^{\rm fin.} =
\left[ 1 - {I}_c \right]  
{\cal M}_{c{\bar c}\to F} \;\;,
\label{mfinall}
\end{equation}
and it can be directly compared with the form of the hard-virtual amplitude 
$\widetilde{\cal M}_{c{\bar c}\to F}$. The relations (\ref{hvall}) and 
(\ref{mfinall}) are in one-to-one correspondence through the simple replacement
$\widetilde{\cal M}_{c{\bar c}\to F} \leftrightarrow 
{\cal M}_{c{\bar c}\to F}^{\rm fin.}$ and 
$\tilde{I}_c \leftrightarrow {I}_c$. Therefore, by requiring that 
$\widetilde{\cal M}_{c{\bar c}\to F}$ is IR finite, we conclude that the 
$\ep$-pole contributions of the operator $\tilde{I}_c(\ep,M^2)$ are exactly the
same as those of ${I}_c$ (equivalently, $\tilde{I}_c$ and ${I}_c$ can differ only
through terms that produce IR finite contributions to 
$\widetilde{\cal M}_{c{\bar c}\to F}$).

We continue our all-order discussion at the fixed-order level to make our
conclusions more explicit and clear. We consider the NLO and NNLO terms. The
extension to higher orders is straightforward.

At the NLO (one-loop) level, $\tilde{I}_a^{(1)}(\ep, M^2/\mu_R^2)$ is written as
in Eq.~(\ref{i1til}) since, as previously discussed, $\tilde{I}_a^{(1)}$
originates from soft terms and from the virtual part of the collinear
counterterm. Both $\tilde{I}_a^{(1) \,{\rm soft}}$ and $\tilde{I}_a^{(1) \,{\rm
coll}}$ contain $\ep$ poles and IR finite contributions. The coefficients of the
$\ep$ poles in Eqs.~(\ref{i1soft}) and (\ref{i1coll})
are determined by the known explicit expression \cite{Catani:1998bh}
of the first-order term  ${I}_a^{(1)}$
of the IR operator ${I}_a$ in Eq.~(\ref{mfinall}).
The dependence on $M^2/\mu_R^2$ is determined to all-order in $\ep$ by
renormalization-group invariance in $d=4-2\ep$ dimensions.
Therefore, the expressions of 
$\tilde{I}_a^{(1) \,{\rm soft}}$ and $\tilde{I}_a^{(1) \,{\rm
coll}}$ in  Eqs.~(\ref{i1soft}) and (\ref{i1coll})
are completely determined apart from an $\ep$-independent contribution
(the inclusion of higher power of $\ep$ is harmless since, in the limit $\ep \to
0$, it gives a vanishing contribution to 
$\widetilde{\cal M}_{c{\bar c}\to F}^{(1)}$ according to Eq.~(\ref{ampli1})).
Then, we note that the entire expression on the right-hand side of 
Eq.~(\ref{i1coll}) {\em exactly} coincides with the virtual part of the collinear
counterterm in the \ms\ factorization scheme \cite{Curci:1980uw}
(we recall that in the \ms\ factorization scheme the collinear counterterm has
only $\ep$-pole contributions, and that the coefficient $\gamma_a$ in
Eqs.~(\ref{i1coll}) and (\ref{galowest})
is equal to the coefficient of the virtual part of the first-order
Altarelli--Parisi splitting function). Therefore, we can conclude that the only
unknown contribution to $\tilde{I}_a^{(1)}$ is an $\ep$-independent term that has
a soft origin. This term is included in the right-hand side of
Eq.~(\ref{i1soft}), and it can be written as $\delta_a = C_a \delta^{q_T}$,
since the intensity of soft radiation from the parton $a$ is simply proportional
to the Casimir coefficient $C_a$ of that parton.
In summary, we have given a proof of the results in 
Eqs.~(\ref{i1til})--(\ref{galowest}), although we cannot give the explicit value
of the process-independent coefficient $\delta^{q_T}$ on the basis of our general
discussion. The explicit determination of $\delta^{q_T}$ requires a detailed
calculation, and the value in Eq.~(\ref{delowest}) is taken from available 
results in the literature \cite{deFlorian:2001zd}.

At the NNLO (two-loop) level we can repeat the same reasoning and steps
as at the NLO (one-loop) level. Using the explicit form of the $\ep$ poles of the
operator ${I}_a^{(2)}$ \cite{Catani:1998bh} in Eq.~(\ref{mfinall})
and the requirement of renormalization-group invariance, we eventually obtain
the second-order subtraction operator $\tilde{I}_a^{(2)}(\ep, M^2/\mu_R^2)$
in the form of Eq.~(\ref{i2til}), where the term 
$\widetilde{H}_a^{(2)} = \widetilde{H}_a^{(2)\,{\rm soft}}
+\widetilde{H}_a^{(2)\,{\rm coll}}$ is completely determined apart from
$\ep$-independent contributions of soft 
and collinear
(from the virtual part of the collinear counterterm) origin. However, the 
$\ep$-independent contribution of collinear origin is vanishing (as in the case
of $\tilde{I}_a^{(1)}$), since 
$\widetilde{H}_a^{(2)\,{\rm coll}}(\epsilon,M^2/\mu_R^2)=
\frac{1}{16 \ep} \gamma_{a (1)} (M^2/\mu_R^2)^{-2\ep}$ {\em exactly} coincides with the
entire collinear-counterterm contribution (in the \ms\ factorization scheme) 
due to the virtual part of the second-order Altarelli--Parisi splitting function
\cite{Curci:1980uw, Furmanski:1980cm}. Therefore, the remaining 
$\ep$-independent contribution to $\widetilde{H}_a^{(2)}$ has a soft origin, and
it can be written in the form $\delta_{a (1)} = C_a \,\delta_{(1)}^{q_T}$ in the
right-hand side of Eq.~(\ref{h2expl}). Owing to its origin from soft
factorization (see the Appendix of Ref.~\cite{Catani:1999ss} and Sect.~5 of
Ref.~\cite{Catani:2000pi}), $\delta_{a (1)}$ is simply proportional to the
Casimir coefficient $C_a$ of the radiating (colliding) parton $a$
and the QCD coefficient $\delta_{(1)}^{q_T}$ is {\em fully} process independent,
namely, it is the {\em same} coefficient for processes that are initiated by either
$q{\bar q}$ annihilation or gluon fusion.

In summary, we have proven the two-loop results in 
Eqs.~(\ref{i2til})--(\ref{donecoef}), although $\delta_{(1)}^{q_T}$ cannot be
determined from our general discussion. The explicit determination of 
$\delta_{(1)}^{q_T}$ requires a detailed calculation. Such a calculation can be
explicitly performed in a general process-independent form by extending
the analysis of Ref.~\cite{deFlorian:2001zd} (which is based on NNLO
soft/collinear factorization formulae \cite{Catani:1999ss, Bern:1999ry,
Catani:2000pi, Campbell:1997hg, Kosower:1999rx}) to the necessary level of
accuracy (in practice, by including contributions at higher order in $\ep$ that
were omitted in the actual computation of Ref.~\cite{deFlorian:2001zd}).
Alternatively, we can exploit our proof of the universality of 
$\delta_{(1)}^{q_T}$ and, therefore, we can determine the value of 
$\delta_{(1)}^{q_T}$ from the NNLO calculation of a single specific process. We
have followed the latter procedure (see below) to obtain the explicit result
of $\delta_{(1)}^{q_T}$ that is reported in Eq.~(\ref{delnnlo}).

The NNLO computation of the DY cross section at small values of $q_T$
was performed in Ref.~\cite{Catani:2009sm},
and the complete result is presented in Ref.~\cite{Catani:2012qa} 
in explicit analytic form.
From Ref.~\cite{Catani:2012qa} we can thus extract the explicit value 
of the NNLO coefficient
$H_q^{DY \,(2)}$ for the DY process (see Appendix~\ref{appa}).
The same value of $H_q^{DY \,(2)}$ is obtained by the fully independent
calculation of Ref.~\cite{Gehrmann:2012ze}.
The scattering amplitude ${\cal M}_{q{\bar q}\to DY}$ for the DY process
was computed long ago up to the two-loop level \cite{dyampli, Matsuura:1988sm}.
Therefore,
using the result of Refs.~\cite{dyampli, Matsuura:1988sm} and the expressions 
in Eqs.~(\ref{ampli0})--(\ref{donecoef}),
we can straightforwardly compute the corresponding
hard-virtual amplitude $\widetilde{\cal M}_{q{\bar q}\to DY}$ and the
corresponding coefficient $H_q^{DY \,(2)}$ from Eq.~(\ref{Hq}). 
Considering the value of $\delta_{(1)}^{q_T}$ as an unknown parameter, we thus
obtain an expression of $H_q^{DY \,(2)}$ (which linearly depends on 
$\delta_{(1)}^{q_T}$) that can be directly compared with the explicit value
extracted from the calculation of Refs.~\cite{Catani:2012qa, Gehrmann:2012ze}. 
This comparison gives the value of $\delta_{(1)}^{q_T}$ that is reported in
Eq.~(\ref{delnnlo}).

The same procedure can be applied to extract the value of $\delta_{(1)}^{q_T}$
from Higgs boson production by gluon fusion. Indeed, also for this process we
know both the explicit value of the coefficient $H_g^{H \,(2)}$ from a direct
NNLO computation \cite{Catani:2011kr} (see Appendix~\ref{appa})
and the corresponding two-loop amplitude
${\cal M}_{gg\to H}$ \cite{ggFF2} (both results use the large-$M_{top}$
approximation). Using these results, we confirm the value of 
$\delta_{(1)}^{q_T}$ that we have extracted from the DY process.

Note that the agreement between these two {\em independent} determinations
(extractions) of $\delta_{(1)}^{q_T}$ is a highly non-trivial check of the
results of Sect.~\ref{sec:hardvirtual}, especially because we are considering two
processes that are controlled by the $q{\bar q}$ annihilation channel and the
gluon fusion channel ($\delta_{(1)}^{q_T}$ is instead independent of the specific
channel). Note also that this agreement can alternatively (i.e., assuming the
knowledge of $\delta_{(1)}^{q_T}$) be regarded as a non-trivial (though partial)
cross-check of the results of the NNLO calculations of DY 
\cite{Catani:2012qa, Gehrmann:2012ze}
and Higgs boson \cite{Catani:2011kr} production.
 
We add a final comment related to our general discussion on the structure of the
hard-virtual term. The all-loop scattering amplitude 
${\cal M}_{c{\bar c}\to F}$ includes an overall phase factor 
$e^{+i \phi_{\rm Coul.}(\ep,M^2)}$ (the phase $\phi_{\rm Coul.}(\ep,M^2)$ is IR
divergent), which is the QCD analogue of the QED Coulomb phase. This phase factor
is physically harmless, since it cancels in the evaluation of the squared
amplitude and, consequently, in the computation of cross sections.
Our explicit expression of $\tilde{I}_c^{(1)}$ includes an imaginary contribution
(the term that is proportional to $i \pi/\ep$ in Eq.~(\ref{i1soft})), and
corresponding contributions are present in the expression 
(\ref{i2til}) of $\tilde{I}_c^{(2)}$ (through its dependence on 
$\tilde{I}_c^{(1)}$). These contributions of `imaginary' origin {\em exactly}
correspond to the perturbative expansion of the Coulomb phase factor, and they
lead to a hard-virtual amplitude $\widetilde{\cal M}_{c{\bar c}\to F}$ that does
not include the overall and harmless (but IR divergent) factor 
$e^{+i \phi_{\rm Coul.}(\ep,M^2)}$. This imaginary contributions to 
$\tilde{I}_c$ cannot arise from a direct computation of $H_c^F$ at the cross
section level and, actually, we have introduced them to the sole practical
(aesthetical) purpose of cancelling the IR divergent Coulomb phase of 
${\cal M}_{c{\bar c}\to F}$. In other words, by removing these contributions
from $\tilde{I}_c^{(1)}$ and $\tilde{I}_c^{(2)}$, we would change the definition
of $\widetilde{\cal M}_{c{\bar c}\to F}$ (by the overall factor 
$e^{+i \phi_{\rm Coul.}(\ep,M^2)}$), but the final results of the hard-virtual
coefficients $H_c^F$ in Eqs.~(\ref{Hq}) and (\ref{Hg}) are unchanged.

\section{Universality and threshold resummation}
\label{sec:thre}

The structure of transverse-momentum resummation and, especially,
of the hard-virtual term can be compared with the analogous structure of
threshold resummation \cite{Sterman:1986aj, Catani:1989ne},
which arises in the context of the QCD computation of the total cross section.
To highlight the main aspects of the comparison, we consider the {\em total}
cross section for the process of Eq.~(\ref{class})
in the simple case (the restriction to this simple case has the sole purpose of
simplifying the notation) in which the final-state system $F$ consists of a
single (`on-shell') particle of mass $M$ (for example, $F$ can be a vector boson
or a Higgs boson). The total cross section $\sigma_F(p_1,p_2;M^2)$
for the production of the system $F$ is computable in QCD perturbation theory
according to the following factorization formula:
\begin{equation}
\sigma_F(p_1,p_2;M^2) = \sum_{a_1, a_2} \int_0^1 dz_1 \int_0^1 dz_2 \;
\;\hat{\sigma}^F_{a_1 a_2}({\hat s}=z_1z_2s;M^2; \as(M^2)) 
\; f_{a_1/h_1}(z_1,M^2)  \; f_{a_2/h_2}(z_2,M^2) 
\;\;,
\label{sigtot}
\end{equation}
where $\hat{\sigma}^F_{a_1 a_2}$ is the total partonic cross section for the
inclusive partonic process $a_1 a_2 \to F + X$ and, for simplicity, the parton
densities $f_{a_i/h_i}(z_i,M^2)$ $(i=1,2)$ are evaluated at the scale $M^2$
(the inclusion of an arbitrary factorization scale $\mu_F$ in the parton
densities and in the partonic cross sections can be implemented in a
straightforward way by using the Altarelli--Parisi evolution equations of
$f_{a/h}(z,\mu_F^2)$).
The partonic cross section
$\hat{\sigma}^F_{a_1 a_2}({\hat s};M^2; \as(M^2))$ depends on the mass $M$
of the system $F$, on the centre--of--mass energy ${\hat s}$ of the colliding
partons, and it is a renormalization-group invariant quantity
that can be perturbatively computed as series expansion in powers of $\as(M^2)$
(equivalently, we can expand $\hat{\sigma}^F_{a_1 a_2}$ in powers of
$\as(\mu_R^2)$, with corresponding perturbative coefficients that explicitly
depend on $M^2/\mu_R^2$).

The kinematical ratio $z=M^2/{\hat s}$ parametrizes the distance from the 
partonic threshold. In the kinematical region close to the partonic threshold
(i.e., where $z \to 1$), the partonic cross section 
$\hat{\sigma}^F_{a_1 a_2}$ receives large QCD radiative corrections of the type
$\left(\f{1}{1-z} \ln^m(1-z)\right)_+$ (the subscript `$+$' denotes the customary
`plus-distribution'). The all-order resummation of these logarithmic contributions
can be systematically performed by working in Mellin ($N$-moment) space
\cite{Sterman:1986aj, Catani:1989ne}. The Mellin transform 
$\hat{\sigma}_{N}(M^2)$ of the partonic cross section 
$\hat{\sigma}({\hat s};M^2)$ is defined as
\begin{equation}
\hat{\sigma}^F_{a_1 a_2, \,N}(M^2; \as(M^2)) \equiv
\int_0^1 dz \; z^{N-1} \;\hat{\sigma}^F_{a_1 a_2}({\hat s}=M^2/z;M^2; \as(M^2))
\;\;.
\end{equation}
In Mellin space, the threshold region $z \to 1$ corresponds to the limit
$N \to \infty$, and the plus-distributions become powers of $\ln N$
( $\left(\f{1}{1-z} \ln^m(1-z)\right)_+ \to \ln^{m+1} N + `{\rm subleading \;
logs}'$). These logarithmic contributions are evaluated to all perturbative 
orders by using threshold resummation \cite{Sterman:1986aj, Catani:1989ne}.
Neglecting terms that are relatively suppressed by powers of $1/N$ in the
limit $N \to \infty$, we write
\begin{equation}
\hat{\sigma}^F_{c {\bar c}, \,N}(M^2; \as(M^2)) =
\hat{\sigma}^{F ({\rm res})}_{c {\bar c}, \,N}(M^2; \as(M^2))
\; \Bigl[ 1 + {\cal O}(1/N) \Bigr] \;\;.
\label{thresfor}
\end{equation}
Note that we are considering only the partonic channel $\,c{\bar c} \to F +X$,
with $c{\bar c}=q{\bar q}$ or $c{\bar c}=gg$, since the other partonic channels
give contributions that are of ${\cal O}(1/N)$.
The expression $\hat{\sigma}^{F ({\rm res})}_{c {\bar c}, \,N}$ in 
the right-hand
side of Eq.~(\ref{thresfor}) embodies all the perturbative terms that are
logarithmically enhanced or {\em constant} in the limit $N \to \infty$.
The partonic cross section $\hat{\sigma}^{F ({\rm res})}_{c {\bar c}, \,N}$
has a universal (process-independent)  all-order structure that is given by the
following threshold-resummation formula
\cite{Sterman:1986aj, Catani:1989ne, Catani:2003zt, Moch:2005ba}:
\begin{equation}
\hat{\sigma}^{F ({\rm res})}_{c {\bar c}, \,N}(M^2; \as(M^2))
= \sigma^{(0)}_{c{\bar c}\to F}(M^2; \as(M^2)) \;
C^{\, \rm th}_{c{\bar c}\to F}(\as(M^2)) \; \Delta_{c, \,N}(M^2) \;\;,
\label{thallorder}
\end{equation}
where $\sigma^{(0)}_{c{\bar c}\to F}$ is the lowest-order cross section
for the partonic process $c{\bar c}\to F$. The radiative factor 
$\Delta_{c, \,N}$ resums all the perturbative contributions $\as^n \ln^m N$
(including some constant terms, i.e. terms with $m=0$). This factor only 
depends 
on the type ($c=q$ or $c=g$) of colliding partons ($\Delta_{c, \,N}$ does not
depend on the final-state system $F$), and it has the form
\begin{equation}
\label{delformfact}
\Delta_{c, \,N}(M^2)= \exp \left\{ \int_{0}^{1} dz \,\frac{z^{N-1}-1}{1-z} 
\left[ 2 \int_{M^2}^{(1-z)^2M^2} \frac{dq^2}{q^2} 
 A_c^{\rm th}(\as(q^2))  + D_c(\as((1-z)^2M^2)) 
 \right] \right\} 
\;\;,
\end{equation}
where $A_c^{\rm th}(\as)$ and $D_c(\as)$ are perturbative series in $\as$,
\begin{equation}
A_c^{\rm th}(\as)=\sum_{n=1}^\infty\left(\frac{\as}{\pi}\right)^n 
A^{{\rm th}\,(n)}_c~~~,~~~~~~~
D_c(\as)=\left(\frac{\as}{\pi}\right)^2 D^{(2)}_c +
\sum_{n=3}^\infty\left(\frac{\as}{\pi}\right)^n D^{(n)}_c\; .
\end{equation}
The perturbative coefficients $A^{{\rm th}\,(1)}_c, A^{{\rm th}\,(2)}_c$
\cite{Catani:1989ne, Catani:1990rr, Catani:1998tm} 
and $A^{{\rm th}\,(3)}_c$ \cite{Moch:2004pa, Moch:2005ba} are explicitly known.
We recall that $A^{{\rm th}\,(1)}_c$ and $A^{{\rm th}\,(2)}_c$ are exactly equal
to the corresponding coefficients $A^{(1)}_c$ and $A^{(2)}_c$
for small-$q_T$ resummation (see Eq.~(\ref{abfun})), while 
$A^{{\rm th}\,(3)}_c \neq A^{(3)}_c $ \cite{Becher:2010tm}.
The perturbative expansion of $D_c(\as)$ starts at ${\cal O}(\as^2)$
(i.e., $D^{(1)}_c=0$), and the perturbative coefficients $D^{(2)}_c$ 
\cite{Vogt:2000ci, Catani:2001ic}
and $D^{(3)}_c$ \cite{Moch:2005ky, Laenen:2005uz} 
are explicitly known.

The factor $C^{\, \rm th}_{c{\bar c}\to F}$ in Eq.~(\ref{thallorder})
embodies remaining $N$-independent contributions (i.e., terms that are constant
in the limit $N \to \infty$) to the partonic cross section. This factor is
definitely process dependent, and it has
the general perturbative expansion
\begin{equation}
C^{\, \rm th}_{c{\bar c}\to F}(\as) = 1 + \sum_{n=1}^\infty
\left(\frac{\as}{\pi}\right)^n C^{\,{\rm th}\,(n)}_{c{\bar c}\to F} \;\;.
\label{cthexp}
\end{equation}
The NLO and NNLO coefficients 
$C^{\,{\rm th}\,(1)}$
and $C^{\,{\rm th}\,(2)}$ are explicitly known in the case of DY
\cite{Matsuura:1988sm, Moch:2005ba}
and Higgs boson
\cite{Kramer:1996iq, Catani:2001ic, Harlander:2001is, Catani:2003zt}
production.
Considering these two specific processes, relations between the $N$-independent
factor $C^{\, \rm th}(\as)$ and the corresponding virtual amplitudes (namely, the
quark and gluon form factors) were discussed and examined in 
Refs.~\cite{Eynck:2003fn, Moch:2005ky, Laenen:2005uz, Ravindran:2006cg, 
Grunberg:2006gd}.
Considering a generic process of the class in Eq.~(\ref{class}) and
using soft-gluon factorization formulae
\cite{Catani:1999ss, Catani:2000pi, Catani:1998bh},
the authors of Ref.~\cite{deFlorian:2012za} have recently shown how the 
NLO and NNLO
coefficients $C^{\,{\rm th}\,(1)}_{c{\bar c}\to F}$ and 
$C^{\,{\rm th}\,(2)}_{c{\bar c}\to F}$ can be directly and explicitly related in
a {\em process-independent} form to the one-loop and two-loop scattering
amplitude ${\cal M}_{c{\bar c}\to F}$
of the underlying partonic process in Eq.~(\ref{partpro}).

The transverse-momentum resummation formula (\ref{qtycross})
has close analogies with the threshold-resummation formula (\ref{thallorder})
(although the latter is somehow simpler).
The process-independent Sudakov form factor $S_c(M,b)$ in Eq.~(\ref{qtycross})
is analogous to the radiative factor $\Delta_{c, \,N}(M^2)$ in 
Eq.~(\ref{thallorder}). Note that $S_c(M,b)$ and $\Delta_{c, \,N}(M^2)$
are both renormalization-group invariant. The analogue of the
process-independent
coefficients $C_{qa}$ and $C^{\mu \nu}_{ga}$ in 
Eqs.~(\ref{qtycross}), (\ref{what}) and (\ref{whatgg}) is absent in the case of
threshold resummation, where there is no ensuing distinction in the resummation
structure between the $q{\bar q}$ annihilation channel and the gluon fusion
channel. The hard-virtual term $H_c^F$ in Eq.~(\ref{qtycross}) is analogous to
the corresponding hard-virtual term $C^{\, \rm th}_{c{\bar c}\to F}$
in Eq.~(\ref{thallorder}).

The analogy between the two hard-virtual terms $H_c^F$ and 
$C^{\, \rm th}_{c{\bar c}\to F}$ can be sharpened. Indeed, we can show that the
{\em all-order} expression of $C^{\, \rm th}_{c{\bar c}\to F}$
can be related to the all-loop scattering amplitude 
${\cal M}_{c{\bar c}\to F}$ of the process in Eq.~(\ref{partpro})
in a process-independent form that is similar to that discussed in 
Sect.~\ref{sec:hardvirtual}. We can write
\begin{equation}
\label{cth}
\as^{2k}(M^2) \;C^{\,\rm th}_{c{\bar c}\to F}(\as(M^2))
=\f{|\widetilde{\cal M}^{\rm th}_{c{\bar c}\to F}|^2}{
|{\cal M}_{c{\bar c}\to F}^{(0)}|^2}\, ,
\end{equation}
with
\begin{equation}
\widetilde{\cal M}^{\rm th}_{c{\bar c}\to F}
= \left[ 1 - \tilde{I}_c^{\,\rm th}(\ep,M^2) \right]  
{\cal M}_{c{\bar c}\to F}
\;\;,
\label{thvall}
\end{equation}
\begin{equation}
\tilde{I}_c^{\,\rm th}(\ep,M^2) = \frac{\as(\mu_R^2)}{2\pi}
\,\tilde{I}_c^{\,{\rm th} (1)}\!\left(\ep,\frac{M^2}{\mu_R^2}\right) + 
\left( \frac{\as(\mu_R^2)}{2\pi} \right)^{\!\!2} 
\tilde{I}_c^{\,{\rm th} (2)}\!\left(\ep,\frac{M^2}{\mu_R^2}\right)
+ \sum_{n=3}^\infty \left( \frac{\as(\mu_R^2)}{2\pi} \right)^{\!\!n}
\tilde{I}_c^{\,{\rm th} (n)}\!\left(\ep,\frac{M^2}{\mu_R^2}\right) \,,
\label{thitilall}
\end{equation}
where $\widetilde{\cal M}^{\rm th}_{c{\bar c}\to F}$ is the (IR finite)
hard-virtual
amplitude for threshold resummation, and the four-dimensional limit 
$\ep \to 0$ is not explicitly denoted in the right-hand side of 
Eq.~(\ref{cth}) (analogously to the case of Eqs.~(\ref{Hq}) and (\ref{Hg})).
The hard-virtual amplitude $\widetilde{\cal M}^{\rm th}_{c{\bar c}\to F}$
is related to the scattering amplitude ${\cal M}_{c{\bar c}\to F}$
by Eq.~(\ref{thvall}), which is completely analogous to Eq.~(\ref{hvall}).
The all-order subtraction operators 
$\tilde{I}_c$ and $\tilde{I}_c^{\,\rm th}$ of Eqs.~(\ref{hvall}) and 
(\ref{thvall}) are {\em different}, since they refer to different physical
observables (namely, the $q_T$-differential cross section versus the total cross
section). Nonetheless, the differences in their structure are minimal.
In particular, $\tilde{I}_c^{(n)}$ 
and 
$\tilde{I}_c^{\,{\rm th} (n)}$, with $n=1,2$~, simply differ by a constant
($\ep$-independent) contribution, namely, the contribution that is parametrized
by the coefficients $\delta^{\,q_T}$ and $\delta_{(1)}^{q_T}$. 
More precisely, the explicit
expression of the NLO term $\tilde{I}_c^{\,{\rm th} (1)}$ is obtained from
Eqs.~(\ref{i1til})--(\ref{galowest})
by simply applying the replacement 
$\tilde{I}_c^{(1)} \to \tilde{I}_c^{\,{\rm th} (1)}$ and 
$\delta^{\,q_T} \to \delta^{\,\rm th}$.  
Then, the explicit
expression of the NNLO term $\tilde{I}_c^{\,{\rm th} (2)}$ is obtained from
Eqs.~(\ref{i2til})--(\ref{donecoef}) by simply applying the replacement 
$\tilde{I}_c^{(2)} \to \tilde{I}_c^{\,{\rm th} (2)}$,
$\tilde{I}_c^{(1)} \to \tilde{I}_c^{\,{\rm th} (1)}$
and 
$\delta_{(1)}^{\,q_T} \to \delta_{(1)}^{\,\rm th}$.
The explicit values of the threshold-resummation coefficients
$\delta^{\,\rm th}$ and $\delta_{(1)}^{\,\rm th}$ are
\begin{equation}
\delta^{\,\rm th}=\delta^{\,q_T} - \,\zeta_2 = - \,\zeta_2 \;\;,
\label{delth}
\end{equation}
\begin{equation}
\delta_{(1)}^{\,\rm th}=\delta_{(1)}^{\,q_T} + \f{40}{3}\zeta_3\pi \beta_0
+ 4 \zeta_2^2 C_A =
\zeta_2\, K+20\,\zeta_3\,\pi \beta_0+C_A\left(-\f{1214}{81}+5\,\zeta_2^2\right)+\f{164}{81}N_f
\;\;.
\label{delth1}
\end{equation}

The results in Eqs.~(\ref{cth})--(\ref{delth1}) are obtained by using the same
reasoning and discussion as in Sect.~\ref{hvstruct} and, in particular,
by exploiting the properties of soft/collinear factorization. We do not repeat
the entire discussion of Sect.~\ref{hvstruct}, and we limit ourselves to
remarking the few points in which the discussion slightly differs. Considering
the computation of the total partonic cross section, we can
directly refer to the classification in contributions from (i) the
elastic-production process, (ii) inelastic processes and (iii) the collinear
counterterm. (i) The elastic process directly contributes to 
$\widetilde{\cal M}^{\,\rm th}_{c{\bar c}\to F}$ in Eq.~(\ref{thvall})
($C^{\,\rm th}_{c{\bar c}\to F}$ in Eq.~(\ref{cth}))
with the all-loop scattering amplitude $\widetilde{\cal M}_{c{\bar c}\to F}$
(with the squared amplitude). (ii) In the kinematical region close to the
partonic threshold, the inelastic processes contribute to the partonic cross
section of Eq.~(\ref{thallorder})
{\em only} through final-state radiation of {\em soft} partons (in the case of
transverse-momentum resummation, collinear radiation also contributes, 
and it is responsible for the presence of the collinear coefficients 
$C_{qa}$ and $C^{\mu \nu}_{ga}$ in Eq.~(\ref{qtycross}) and for the differences 
between the $q{\bar q}$ and $gg$ channels). Soft factorization at the (squared)
amplitude level is not spoiled by kinematical effects, since the kinematics of
the total cross section is exactly factorized 
\cite{Sterman:1986aj, Catani:1989ne}
by the Mellin transformation to $N$ space. This leads to the same conclusion as
in Sect.~\ref{hvstruct} about the contribution of the inelastic processes to the
hard-virtual term. This contribution is factorized and it has a soft origin in
both cases of transverse-momentum and threshold resummation. In particular,
at the cross section level (i.e., after the corresponding phase space
integration), this soft term produces contributions to the coefficients   
$\delta^{\,\rm th}$ and $\delta_{(1)}^{\,\rm th}$ in the expressions of 
$\tilde{I}_c^{\,\rm th}$ and $C^{\,\rm th}_{c{\bar c}\to F}$,
analogously to the corresponding contributions to 
the coefficients   
$\delta^{\,q_T}$ and $\delta_{(1)}^{q_T}$ in the expressions of 
$\tilde{I}_c$ and $H_c^F$. (iii) The radiative factor $\Delta_{c, \,N}$ in
Eq.~(\ref{thallorder})
is entirely due to soft radiation \cite{Sterman:1986aj, Catani:1989ne}.
Therefore,
the complete virtual part of the collinear counterterm in the \ms\ factorization
scheme directly contributes to 
$\widetilde{\cal M}^{\,\rm th}_{c{\bar c}\to F}$ and 
$C^{\,\rm th}_{c{\bar c}\to F}$ (analogously to the contribution to 
$\widetilde{\cal M}_{c{\bar c}\to F}$ and $H_c^F$). It follows that the
collinear counterterm contributions to
$\tilde{I}_c^{\,\rm th}$ and $\tilde{I}_c$ are completely analogous.

Owing to their process independence, the threshold resummation coefficients
$\delta^{\,\rm th}$ and $\delta_{(1)}^{\,\rm th}$ (analogously to 
$\delta^{\,q_T}$ and $\delta_{(1)}^{q_T}$) can be explicitly evaluated from
either the NNLO calculation of a single process or an NNLO calculation in a
process-independent form. We have explicitly verified that the general results
in Eqs.~(\ref{cth})--(\ref{thitilall}) and the explicit values of 
the coefficients in Eqs.~(\ref{delth}) and (\ref{delth1})
are consistent with the NNLO results of the process-independent calculation
of Ref.~\cite{deFlorian:2012za}.

The close correspondence between the hard-virtual terms $H_c^F$ and 
$C^{\,\rm th}_{c{\bar c}\to F}$ of the resummation formulae in 
Eqs.~(\ref{qtycross}) and (\ref{thallorder})
can also be expressed in {\em direct} form. 
Using Eqs.~(\ref{Hq})--(\ref{Hgtens}), 
Eq.~(\ref{cth})
and the expressions (\ref{hvall}) and (\ref{thvall})
of the corresponding hard-virtual amplitudes, we obtain
\begin{align}
\label{hvratio}
\frac{H_c^F(\as)}{C^{\,\rm th}_{c{\bar c}\to F}(\as)}
&=\left\{
\f{|1 - \tilde{I}_c(\ep,M^2)|^2}{
|1 - \tilde{I}_c^{\,\rm th}(\ep,M^2)|^2}
\right\}_{\ep=0} \\
&=
\exp\left\{ \f{\as}{\pi} C_c \left(\delta^{\,q_T} - \delta^{\,\rm th} \right)
+ \left(\f{\as}{\pi}\right)^2 C_c 
\left[ \f{1}{2} K \left(\delta^{\,q_T} - \delta^{\,\rm th} \right)
- \f{1}{8} \left(\delta_{(1)}^{\,q_T} - \delta_{(1)}^{\,\rm th} \right)\right]
+ {\cal O}(\as^3) \right\} 
\label{expratio} \\
&=\exp\left\{ \f{\as}{\pi} C_c \,\zeta_2 
+ \left(\f{\as}{\pi}\right)^2 C_c 
\left[ \f{5}{3} \,\zeta_3 \pi \beta_0
+ \zeta_2 \left( \f{67}{36} C_A - \f{5}{18} N_f \right)\right]
+ {\cal O}(\as^3)
\right\}\, , \nn
\end{align}
where, in the case of gluon fusion processes, the numerator 
in the left-hand side of Eq.~(\ref{hvratio}) is defined as 
$H_g^F \equiv g_{\mu_1 \nu_1} g_{\mu_2 \nu_2} H_g^{F \,\mu_1 \nu_1, \mu_2
\nu_2}$. The equality in Eq.~(\ref{expratio}) is obtained by using the explicit
expression of $\tilde{I}_c(\ep,M^2)$ (see Eqs.~(\ref{i1til})--(\ref{donecoef}))
and the corresponding expression of $\tilde{I}_c^{\,\rm th}(\ep,M^2)$
(see Eq.~(\ref{thitilall}) and accompanying comments)
up to the NNLO.

Considering the ratio of hard-virtual terms for a specific process as in 
Eq.~(\ref{hvratio}), the effect of the all-loop amplitude 
${\cal M}_{c{\bar c}\to F}$ (and the associated process dependence) entirely
cancels. This ratio is completely determined by the contribution of the
inelastic processes (namely, the factorized radiation of final-state partons and
the corresponding virtual corrections) to the corresponding cross sections. The
ensuing IR (soft and collinear) singularities completely cancel, and the final
expression in Eq.~(\ref{expratio}) is entirely determined by the IR finite
contributions due to the (real) emission of {\em soft} QCD radiation.
The {\em exponent} in Eq.~(\ref{expratio}) is directly proportional to the
Casimir factor $C_c$ (i.e., the colour charge) of the colliding partons $c$ and
${\bar c}\;$: this proportionality is a straightforward consequence of the
exponentiating correlation structure of the factorization formulae for {\em
soft-parton radiation} from QCD squared amplitudes 
\cite{Catani:1999ss, Catani:2000pi}.
The value of the perturbative coefficients $\delta^{\,q_T} - \delta^{\,\rm th}$
and $\delta_{(1)}^{\,q_T} - \delta_{(1)}^{\,\rm th}$ in the exponent has a
kinematical origin.

\section{Summary}
\label{sec:summa}

In this paper we have considered QCD radiative corrections to the production
of a generic colourless high-mass system $F$ in hadronic collisions 
(see Eq.~(\ref{class})). 
Large logarithmic terms arise in the QCD perturbative expansion 
when the high-mass system $F$ is produced at small transverse momentum.
These logarithmic terms can be resummed to all perturbative orders by using a
universal (process-independent) resummation formula (see Sect.~\ref{sec:resu}) 
and, then,
they are controlled by a set of resummation factors and ensuing
perturbative resummation coefficients.
After having recalled the process independence of the Sudakov form factor 
and the
explicit expressions (up to NNLO) of the process-independent
{\em collinear} coefficients (see Sect.~\ref{sec:procind}), 
in Sect.~\ref{sec:hardvirtual}
we have focused on the hard-virtual factor
$H_c^F$.
We have shown that, although this 
factor
is process dependent, 
it can be directly related 
(see Eqs.~(\ref{Hq})--(\ref{Hgtens}))
in a universal (process-independent) way to 
the IR finite part of the all-order virtual amplitude
${\cal M}_{c{\bar c}\to F}$ of the corresponding partonic subprocess 
$c{\bar c}\to F$.
Therefore, the all-order scattering amplitude ${\cal M}_{c{\bar c}\to F}$
is the {\em sole} process-dependent information that is eventually required
by the all-order resummation formula. 
The relation between $H_c^F$ and ${\cal M}_{c{\bar c}\to F}$
follows from a universal all-order factorization formula 
(see Eqs.~(\ref{hvall}) and (\ref{itilall})) that originates from
the factorization properties of soft (and collinear) parton radiation.
We have explicitly determined this relation up to the NNLO.
More precisely, we have shown that 
this relation is fully determined by the structure of IR singularities of the
all-order amplitude ${\cal M}_{c{\bar c}\to F}$ and by
renormalization-group invariance up to a {\em single} coefficient (of
{\em soft} origin) at each
perturbative order.
We have explicitly determined these coefficients at NLO and NNLO.
Therefore, knowing the NNLO scattering amplitude ${\cal M}_{c{\bar c}\to F}$,
its corresponding hard-virtual resummation factor $H_c^F$ is straightforwardly
determined up to NNLO.

The results presented in this paper, with the knowledge of the other 
process-independent resummation coefficients (which are recalled in 
Sects.~\ref{sec:resu} and \ref{sec:procind}), complete 
(modulo the second-order coefficients $G_{ga}^{(2)}$, as discussed at the end of
Sect.~\ref{sec:procind})
the $q_T$ resummation
formalism in explicit form up to full NNLL and NNLO accuracy
for all the processes in the class of Eq.~(\ref{class}).
This permits applications to NNLL+NNLO resummed calculations 
for {\em any} processes whose NNLO scattering amplitudes are available.
Moreover, since the hard-virtual (and collinear) resummation coefficients
are exactly the coefficients that are required to implement the 
$q_T$ subtraction formalism \cite{Catani:2007vq}, the results that we have
presented
are directly and straightforwardly applicable to perform fully-exclusive
NNLO computations for each of these processes.

We have also considered the related process-independent formalism of threshold
resummation (see Sect.~\ref{sec:thre}). We have shown that the process-dependent
hard-virtual factor $C^{\,\rm th}_{c{\bar c}\to F}$ of threshold resummation
has a universal (process-independent) structure that is analogous to that of the 
hard-virtual factor $H_c^F$ of transverse-momentum resummation. The 
process-independent relation between $C^{\,\rm th}_{c{\bar c}\to F}$
and the scattering amplitude ${\cal M}_{c{\bar c}\to F}$ has been explicitly
pointed out up to NNLO. In particular, we have shown that, for each specific
process, the ratio $H_c^F/C^{\,\rm th}_{c{\bar c}\to F}$ is completely
independent of the process (i.e., independent of ${\cal M}_{c{\bar c}\to F}$),
and it is fully determined by the associated {\em soft-parton} radiation.

\noindent {\bf Acknowledgements.}
This work was supported in part by UBACYT, CONICET, ANPCyT, INFN and
the Research Executive Agency (REA) 
of the European Union under the Grant Agreement number PITN-GA-2010-264564 
({\it LHCPhenoNet},  Initial Training Network).

\appendix

\section{Hard-virtual coefficients in DY, Higgs boson and diphoton production}
\label{appa}

In this Appendix we report the explicit expressions of the hard-virtual coefficients
in the hard scheme for the cases of DY, Higgs boson and diphoton production.
The NNLO coefficients  $H_q^{DY(2)}$ and $H_g^{H(2)}$ for DY and Higgs boson
(using the large-$m_Q$ approximation) production were obtained in 
Refs.~\cite{Catani:2012qa} and \cite{Catani:2011kr} by performing a direct QCD
computation of the corresponding $q_T$ cross sections. The same results can be
recovered (as discussed in Sect.~\ref{hvstruct})
by using the process-independent structure of the hard-virtual coefficients.
The explicit expression (which is presented in Eq.~(\ref{h22gamma}))
of the NNLO coefficient $H_q^{\gamma \gamma (2)}$ for diphoton production is
directly obtained by using the results of Sect.~\ref{sec:hardvirtual}.

Starting from the DY process, we consider the production (through the $q{\bar q}$
annihilation channel) of a virtual photon or a vector boson ($V=\gamma^*,W^\pm,Z$)
and the subsequent leptonic decay.
The corresponding Born-level cross section 
$\left[d\sigma_{q{\bar q},V}^{(0)}\right]$ in Eq.~(\ref{qtycross})
depends on the kinematics of the leptonic final state, while the hard-virtual
term $H_q^{DY}$ only depends on the partonic process $q{\bar q} \to V$.
In particular, $H_q^{DY}(\as(M^2))$ only depends on $\as(M^2)$, with no additional
dependence on kinematical variables.
The NLO and NNLO hard-virtual coefficients $H_q^{DY(1)}$ and $H_q^{DY(2)}$
in the hard scheme are \cite{Catani:2012qa}
\begin{equation}
\label{H1q}
H_q^{DY(1)}=C_F\left(\f{\pi^2}{2}-4\right) \;\;,
\end{equation}
\begin{align}
\label{H2q}
H_q^{DY(2)}&=C_FC_A\left(\frac{59 \zeta_3}{18}-\frac{1535}{192}+\frac{215 \pi ^2}{216}-\frac{\pi ^4}{240}\right)
+\frac{1}{4} C_F^2\left(-15 \zeta_3+\frac{511}{16}-\frac{67 \pi ^2}{12}+\frac{17 \pi ^4}{45}\right)\nn\\
&+\frac{1}{864} C_FN_f  \left(192 \zeta_3+1143-152 \pi ^2\right)\, .
\end{align}

We then consider the production of the SM Higgs boson $H$
through the gluon fusion channel, where $H$ couples to a heavy-quark loop.
If the Higgs boson decays non-hadronically (e.g., 
$H \to ZZ \to 4 \, {\rm leptons}$,
or $H \to \gamma \gamma$), the dependence on the kinematics of its decay products
only affects the 
Born-level cross section 
$\left[d\sigma_{gg,H}^{(0)}\right]$ in Eq.~(\ref{qtycross}), while the
corresponding hard-virtual factor only depends on $\as(M^2)$ (as in the case of DY
production) and on the mass of the heavy quark in the loop. The same conclusion
applies to the hadronic decay of $H$, if we neglect the QCD interferences between
the initial and final states. In both cases (i.e., non-hadronic decay or hadronic
decay without interferences), the spin (Lorentz index) correlation structure of
the hard-collinear factor in Eq.~(\ref{whatgg}) can be simplified. Indeed, as
shown in Ref.~\cite{Catani:2010pd}, the right-hand side of Eq.~(\ref{whatgg})
turns out to be proportional to the following (Lorentz scalar)
hard-virtual factor:
\begin{equation}
\label{hgH}
H_g^H(\as(M^2)) = \;g_{\mu_1 \nu_1} \;g_{\mu_2 \nu_2}
\;H_{g}^{H\,\mu_1\nu_1,\mu_2\nu_2}(\as(M^2))  \;\;,
\end{equation}
whose perturbative coefficients $H_g^{H \,(n)}$ follow from the perturbative
expansion in Eq.~(\ref{hexpgg}).
Assuming that the Higgs boson couples to a single heavy quark of mass $m_Q$,
the first-order coefficient $H_g^{H(1)}$ in the hard scheme is
\begin{equation}
\label{H1g}
H_g^{H(1)}=C_A\pi^2/2+c_H(m_Q)\, .
\end{equation}
The expression in Eq.~(\ref{H1g}) is obtained by using the process-independent
formulae in Eqs.~(\ref{ampli1}), (\ref{Hg}) and (\ref{Hgtens}),
and the function $c_H(m_Q)$ depends on the NLO corrections of the scattering
amplitude  ${\cal M}_{gg \to H}$.
The 
function $c_H(m_Q)$ is given in Eq.~(B.2) of Ref.~\cite{Spira:1995rr} in terms of
one-dimensional integrals; analytic expressions of $c_H(m_Q)$ 
in terms of harmonic polylogarithms are given 
in Eq.~(3.5) of Ref.~\cite{Harlander:2005rq} and Eq.~(27) 
of Ref.~\cite{Aglietti:2006tp}.
In the limit $m_Q\to \infty$, the function $c_H$ becomes
\begin{equation}
c_H(m_Q)\longrightarrow\f{5C_A-3C_F}{2}=\f{11}{2}\, .
\end{equation}
In the large-$m_Q$ limit, the explicit expression of the NNLO hard-virtual
coefficient $H_g^{H(2)}$ in the hard scheme is \cite{Catani:2011kr}
\begin{align}
\label{H2g}
H_g^{H(2)}&=C_A^2\left(
\f{3187}{288}+\f{7}{8}L_Q+\f{157}{72}\pi^2+\f{13}{144}\pi^4-\f{55}{18}\zeta_3\right)+C_A\, C_F\left(-\f{145}{24}-\f{11}{8}L_Q-\f{3}{4}\pi^2\right)\nn\\
&+\f{9}{4}C_F^2
-\f{5}{96}C_A-\f{1}{12}C_F-C_A\, N_f\left(\f{287}{144}+\f{5}{36}\pi^2+\f{4}{9}\zeta_3\right)
+C_F\, N_f\left(-\f{41}{24}+\f{1}{2}L_Q+\zeta_3\right)\, ,
\end{align}
where $L_Q=\ln (M^2/m_Q^2)$.
The scattering amplitude ${\cal M}_{gg \to H}$ has been computed 
\cite{Harlander:2009bw} up to its NNLO by including corrections to the 
large-$m_Q$ approximation (the evaluation of the corrections uses the expansion
parameter $1/m_Q^2$). Using the process-independent
formulae in Eqs.~(\ref{ampli2}), (\ref{Hg}) and (\ref{Hgtens}), 
these corrections can be straightforwardly included in the expression
of the NNLO hard-virtual
coefficient $H_g^{H(2)}$.

We finally consider the case of diphoton production. In this case, the production
cross section receives contributions from both the $q{\bar q}$ annihilation and
gluon fusion channels. The final-state system $F=\gamma \gamma$ can be produced by
the elastic partonic subprocesses $q{\bar q} \to \gamma \gamma$ 
and $g g \to \gamma \gamma$. In the perturbative evaluation of the cross section,
the subprocess $q{\bar q} \to \gamma \gamma$ first contributes at the LO (through
the corresponding tree-level scattering amplitude), while the subprocess
$g g \to \gamma \gamma$ starts to contributes at the NNLO (through the
corresponding scattering amplitude that involves the one-loop QCD interaction of
light and heavy quarks). Therefore, to the purpose of evaluating the complete $q_T$
cross section up to its NNLO, the gluon fusion channel only contributes at its
corresponding lowest order. Having computed
$\left[d\sigma_{gg,\gamma \gamma}^{(0)}\right]$, the NNLO contribution to
Eq.~(\ref{qtycross}) from the gluon fusion channel is obtained by simply
considering the lowest-order hard-virtual coefficient $H_{g}^{\gamma \gamma(0)}$
(in practice, we can simply implement the replacement 
$H_{g}^{\gamma \gamma \,\mu_1\nu_1,\mu_2\nu_2} \to
H_{g}^{\gamma \gamma(0)\mu_1\nu_1,\mu_2\nu_2} \to d^{\mu_1\nu_1}(p_1,p_2) \,
d^{\mu_2\nu_2}(p_1,p_2)/4$ in Eqs.~(\ref{whatgg}) and (\ref{hexpgg})).
The hard-virtual coefficient $H_{q}^{\gamma \gamma}$ of the subprocess 
$q{\bar q} \to \gamma \gamma$ has instead to be explicitly evaluated up to its
NNLO (i.e., we need the perturbative coefficients 
$H_{q}^{\gamma \gamma(1)}$ and $H_{q}^{\gamma \gamma(2)}$).

Using the notation of Eq.~(\ref{partpro}), to compute $H_{q}^{\gamma \gamma}$
we have to consider the partonic process 
$q({\hat p}_1)\bar q({\hat p}_2)\to \gamma(q_1)\gamma(q_2)$, whose Mandelstam 
kinematical variables are 
\begin{equation}
{\hat s}=({\hat p}_1+{\hat p}_2)^2= M^2~~,~~~{\hat u}=({\hat
p}_2-q_1)^2~~,~~~~{\hat t}=({\hat p}_1-q_2)^2\, ,
\end{equation}
with the constraint ${\hat s}+{\hat t}+{\hat u}=M^2+{\hat t}+{\hat u}=0$.
Unlike the case of DY and Higgs boson production, the hard-virtual term 
$H_q^{\gamma\gamma}$ of Eq.~(\ref{what}) depends on the kinematical variables
of the final-state diphoton system. We explicitly specify these two variables
(they are generically denoted as ${\bf\Omega} = \{\Omega_A,\Omega_B \}$
in Eqs.~(\ref{diffxs}), (\ref{qtycross}) and (\ref{what})) by using the azimuthal
angle of $q_1$ and the ratio $v=-{\hat u}/{\hat s}= -{\hat u}/M^2$.
The hard-virtual term $H_q^{\gamma\gamma}$ is invariant with respect to 
azimuthal rotations (see Eq.~(\ref{Hq})) and it is a dimensionless function of
its kinematical variables. This implies that $H_q^{\gamma\gamma}$
only depends on $v$, and we simply use the notation 
$H_q^{\gamma\gamma}(v;\as(M^2))$. The first-order coefficient 
$H_q^{\gamma\gamma(1)}(v)$ of Eq.~(\ref{hexp}) is known \cite{Balazs:1997hv}.
Its explicit expression in the hard scheme is
\begin{align}
H_q^{\gamma\gamma(1)}(v)&=\frac{C_F}{2}
\bigg\{ (\pi^2 -7)
   +\frac{1}{(1-v)^2+ v^2} 
\bigg[\left((1-v)^2+1\right) \ln^2(1-v)+v (v+2)
   \ln (1-v)\nn\\
&+\left(v^2+1\right) \ln^2 v +(1-v)(3-v) \ln v
\bigg]\bigg\} \;\;.
\end{align}
The second-order coefficient $H_q^{\gamma\gamma(2)}(v)$ 
was computed and used in Ref.~\cite{Catani:2011qz}.
The calculation of $H_q^{\gamma\gamma(2)}$ is performed by using the universality
structure of the hard-virtual term 
(see Eqs.~(\ref{ampli0})--(\ref{ampli2}) and (\ref{Hq}))
and the explicit result \cite{Anastasiou:2002zn}
of the two-loop amplitude
${\cal M}_{q{\bar q} \to \gamma \gamma}$ of the process
$q{\bar q} \to \gamma \gamma$\footnote{We note that there are some typos in
Ref.~\cite{Anastasiou:2002zn}: in Eq.~(3.13) the factor 
${\Gamma(1-\epsilon)}/{\Gamma(1-2\epsilon)}$ has to be
replaced with the factor
${\Gamma(1-2\epsilon)}/{\Gamma(1-\epsilon)}$;
the overall sign on the right-hand side of Eqs.~(C.1), (C.2) and (C.3)
has to be reversed (in particular, without this flip of sign, the 
coefficients
of the IR poles of the NLO and NNLO scattering amplitude have wrong 
signs).}.
The result that we obtain for $H_q^{\gamma\gamma(2)}$ in the hard scheme is
\begin{align}
\label{h22gamma}
H_q^{\gamma\gamma(2)}(v)&=\frac{1}{4{\cal A}_{LO}(v)}\left[{\cal F}_{inite,q{\bar q}\gamma\gamma;s}^{0\times 2}+{\cal F}_{inite,q{\bar q}\gamma\gamma;s}^{1\times 1}\right]+3\zeta_2\,C_F H_q^{\gamma\gamma(1)}(v)\nn\\
&-\f{45}{4}\zeta_4\, C_F^2
+C_FC_A\left(\f{607}{324}+\f{1181}{144}\zeta_2-\f{187}{144}\zeta_3-\f{105}{32}\zeta_4\right)\nn\\
&+C_FN_f\left(-\f{41}{162}-\f{97}{72}\zeta_2+\f{17}{72}\zeta_3\right)\, ,
\end{align}
where the functions ${\cal F}_{inite,q{\bar q}\gamma\gamma;s}^{0\times 2}$ and
${\cal F}_{inite,q{\bar q}\gamma\gamma;s}^{1\times 1}$ are defined  
in Eqs.~(4.6) and (5.3) of Ref.~\cite{Anastasiou:2002zn}, respectively, and the
function ${\cal A}_{LO}(v)$ is
\begin{equation}
{\cal A}_{LO}(v)=8\,N_c\f{1-2v+2v^2}{v(1-v)}\, .
\end{equation}


\begin{thebibliography}{99}


\bibitem{Dokshitzer:hw}
Y.~L.~Dokshitzer, D.~Diakonov and S.~I.~Troian,
Phys.\ Lett.\  B {\bf 79} (1978) 269,
Phys.\ Rep.\  {\bf 58} (1980) 269.


\bibitem{Parisi:1979se}
G.~Parisi and R.~Petronzio,
Nucl.\ Phys.\ B {\bf 154} (1979) 427.


\bibitem{Curci:1979bg}
G.~Curci, M.~Greco and Y.~Srivastava,
Nucl.\ Phys.\ B {\bf 159} (1979) 451.

\bibitem{Collins:1981uk}
J.~C.~Collins and D.~E.~Soper,
Nucl.\ Phys.\ B {\bf 193} (1981) 381
[Erratum-ibid.\ B {\bf 213} (1983) 545],
Nucl.\ Phys.\ B {\bf 197} (1982) 446.


\bibitem{Kodaira:1981nh}
J.~Kodaira and L.~Trentadue,
Phys.\ Lett.\ B {\bf 112} (1982) 66,
report SLAC-PUB-2934 (1982),
Phys.\ Lett.\ B {\bf 123} (1983) 335.


\bibitem{Collins:1984kg}
J.~C.~Collins, D.~E.~Soper and G.~Sterman,
Nucl.\ Phys.\ B {\bf 250} (1985) 199.

\bibitem{Catani:vd}
S.~Catani, E.~D'Emilio and L.~Trentadue,
Phys.\ Lett.\ B {\bf 211} (1988) 335.

\bibitem{deFlorian:2000pr}
  D.~de Florian and M.~Grazzini,
  Phys.\ Rev.\ Lett.\  {\bf 85} (2000) 4678
[arXiv:hep-ph/0008152].


\bibitem{Catani:2000vq}
  S.~Catani, D.~de Florian and M.~Grazzini,
  Nucl.\ Phys.\  B {\bf 596} (2001) 299
[arXiv:hep-ph/0008184].


\bibitem{Catani:2010pd}
  S.~Catani and M.~Grazzini,
  Nucl.\ Phys.\ B {\bf 845} (2011) 297
[arXiv:1011.3918 [hep-ph]].


\bibitem{Zhu:2012ts}
  H.~X.~Zhu, C.~S.~Li, H.~T.~Li, D.~Y.~Shao and L.~L.~Yang,
  Phys.\ Rev.\ Lett.\  {\bf 110} (2013) 082001
  [arXiv:1208.5774 [hep-ph]],
  Phys.\ Rev.\ D {\bf 88} (2013) 074004
  [arXiv:1307.2464 [hep-ph]].




\bibitem{Davies:1984hs}
C.~T.~H.~Davies and W.~J.~Stirling,
Nucl.\ Phys.\  B {\bf 244} (1984) 337;
  C.~T.~H.~Davies, B.~R.~Webber and W.~J.~Stirling,
  Nucl.\ Phys.\  B {\bf 256} (1985) 413.


\bibitem{deFlorian:2001zd}
  D.~de Florian and M.~Grazzini,
  Nucl.\ Phys.\  B {\bf 616} (2001) 247
[arXiv:hep-ph/0108273].


\bibitem{Becher:2010tm}
  T.~Becher, M.~Neubert,
  Eur.\ Phys.\ J.\  {\bf C71 } (2011)  1665
[arXiv:1007.4005 [hep-ph]].


\bibitem{Catani:2011kr}
  S.~Catani and M.~Grazzini,
  Eur.\ Phys.\ J.\ C {\bf 72} (2012) 2013
[Erratum-ibid.\ C {\bf 72} (2012) 2132]
[arXiv:1106.4652 [hep-ph]].

\bibitem{Catani:2012qa}
  S.~Catani, L.~Cieri, D.~de Florian, G.~Ferrera and M.~Grazzini,
  Eur.\ Phys.\ J.\ C {\bf 72} (2012) 2195
  [arXiv:1209.0158 [hep-ph]].

 
  

\bibitem{Gao:2005iu}
  Y.~Gao, C.~S.~Li and J.~J.~Liu,
  Phys.\ Rev.\ D {\bf 72} (2005) 114020
  [hep-ph/0501229];
  A.~Idilbi, X.~-d.~Ji and F.~Yuan,
  Phys.\ Lett.\ B {\bf 625} (2005) 253
  [hep-ph/0507196].
\bibitem{Mantry:2009qz}
  S.~Mantry and F.~Petriello,
  Phys.\ Rev.\ D {\bf 81} (2010) 093007
  [arXiv:0911.4135 [hep-ph]],
  Phys.\ Rev.\ D {\bf 83} (2011) 053007
  [arXiv:1007.3773 [hep-ph]].

\bibitem{GarciaEchevarria:2011rb}
  M.~G.~Echevarria, A.~Idilbi and I.~Scimemi,
  JHEP {\bf 1207} (2012) 002
  [arXiv:1111.4996 [hep-ph]];
  M.~G.~Echevarria, A.~Idilbi, A.~Sch\"afer and I.~Scimemi,
  Eur.\ Phys.\ J.\ C {\bf 73} (2013) 2636
  [arXiv:1208.1281 [hep-ph]].

\bibitem{Chiu:2012ir}
  J.~-Y.~Chiu, A.~Jain, D.~Neill and I.~Z.~Rothstein,
  JHEP {\bf 1205} (2012) 084
  [arXiv:1202.0814 [hep-ph]].

\bibitem{Collins:2012uy}
  J.~C.~Collins and T.~C.~Rogers,
  Phys.\ Rev.\ D {\bf 87} (2013) 3,  034018
  [arXiv:1210.2100 [hep-ph]];
  M.~G.~Echevarria, A.~Idilbi and I.~Scimemi,
  Phys.\ Lett.\ B {\bf 726} (2013) 795
  [arXiv:1211.1947 [hep-ph]].

\bibitem{Becher:2012yn}
  T.~Becher, M.~Neubert and D.~Wilhelm,
  JHEP {\bf 1305} (2013) 110
  [arXiv:1212.2621 [hep-ph]].


\bibitem{Gehrmann:2012ze}
  T.~Gehrmann, T.~Lubbert and L.~L.~Yang,
  Phys.\ Rev.\ Lett.\  {\bf 109} (2012) 242003
  [arXiv:1209.0682 [hep-ph]].



\bibitem{Catani:1990xk}
  S.~Catani, M.~Ciafaloni and F.~Hautmann,
  Phys.\ Lett.\ B {\bf 242} (1990) 97,
  Nucl.\ Phys.\ B {\bf 366} (1991) 135;
  J.~C.~Collins and R.~K.~Ellis,
  Nucl.\ Phys.\ B {\bf 360} (1991) 3;
  E.~M.~Levin, M.~G.~Ryskin, Y.~.M.~Shabelski and A.~G.~Shuvaev,
  Sov.\ J.\ Nucl.\ Phys.\  {\bf 53} (1991) 657
   [Yad.\ Fiz.\  {\bf 53} (1991) 1059];
  M.~Deak, F.~Hautmann, H.~Jung and K.~Kutak,
  JHEP {\bf 0909} (2009) 121
  [arXiv:0908.0538 [hep-ph]].

\bibitem{D'Alesio:2007jt}
  U.~D'Alesio and F.~Murgia,
  Prog.\ Part.\ Nucl.\ Phys.\  {\bf 61} (2008) 394
  [arXiv:0712.4328 [hep-ph]].

\bibitem{Rogers:2010dm}
  T.~C.~Rogers and P.~J.~Mulders,
  Phys.\ Rev.\ D {\bf 81} (2010) 094006

\bibitem{Collins:2011zzd}
J.~Collins, {\it Foundations of Perturbative QCD} (Cambridge University Press,
Cambridge, 2011).


\bibitem{Catani:1998bh}
  S.~Catani,
  Phys.\ Lett.\ B {\bf 427} (1998) 161
  [hep-ph/9802439].



\bibitem{Sterman:1986aj}
  G.~F.~Sterman,
  Nucl.\ Phys.\ B {\bf 281} (1987) 310.

\bibitem{Catani:1989ne}
  S.~Catani and L.~Trentadue,
  Nucl.\ Phys.\ B {\bf 327} (1989) 323,
  Nucl.\ Phys.\ B {\bf 353} (1991) 183.


\bibitem{Bozzi:2005wk}
  G.~Bozzi, S.~Catani, D.~de Florian and M.~Grazzini,
  Nucl.\ Phys.\ B {\bf 737} (2006) 73
[arXiv:hep-ph/0508068].

\bibitem{deFlorian:2011xf}
  D.~de Florian, G.~Ferrera, M.~Grazzini and D.~Tommasini,
  JHEP {\bf 1111} (2011) 064
  [arXiv:1109.2109 [hep-ph]],
  JHEP {\bf 1206} (2012) 132
  [arXiv:1203.6321 [hep-ph]].

\bibitem{Wang:2012xs}
  J.~Wang, C.~S.~Li, Z.~Li, C.~P.~Yuan and H.~T.~Li,
  Phys.\ Rev.\ D {\bf 86} (2012) 094026
  [arXiv:1205.4311 [hep-ph]].

\bibitem{Bozzi:2010xn}
  G.~Bozzi, S.~Catani, G.~Ferrera, D.~de Florian and M.~Grazzini,
  Phys.\ Lett.\ B {\bf 696} (2011) 207
  [arXiv:1007.2351 [hep-ph]].

\bibitem{Guzzi:2013aja}
  M.~Guzzi, P.~M.~Nadolsky and B.~Wang,
  arXiv:1309.1393 [hep-ph].



\bibitem{Catani:2007vq}
  S.~Catani and M.~Grazzini,
  Phys.\ Rev.\ Lett.\  {\bf 98} (2007) 222002
[arXiv:hep-ph/0703012].

\bibitem{Grazzini:2008tf}
  M.~Grazzini,
  JHEP {\bf 0802} (2008) 043
[arXiv:0801.3232 [hep-ph]].

\bibitem{Catani:2009sm}
  S.~Catani, L.~Cieri, G.~Ferrera, D.~de Florian and M.~Grazzini,
  Phys.\ Rev.\ Lett.\  {\bf 103} (2009) 082001
[arXiv:0903.2120 [hep-ph]].

\bibitem{Ferrera:2011bk}
  G.~Ferrera, M.~Grazzini and F.~Tramontano,
  Phys.\ Rev.\ Lett.\  {\bf 107} (2011) 152003
[arXiv:1107.1164 [hep-ph]].

\bibitem{Catani:2011qz}
  S.~Catani, L.~Cieri, D.~de Florian, G.~Ferrera and M.~Grazzini,
  Phys.\ Rev.\ Lett.\  {\bf 108} (2012) 072001
  [arXiv:1110.2375 [hep-ph]].
 
\bibitem{Grazzini:2013bna}
  M.~Grazzini, S.~Kallweit, D.~Rathlev and A.~Torre,
  report ZU-TH-21-13 (arXiv:1309.7000 [hep-ph]).


\bibitem{Nadolsky:2007ba}
  P.~M.~Nadolsky, C.~Balazs, E.~L.~Berger and C.~-P.~Yuan,
  Phys.\ Rev.\ D {\bf 76} (2007) 013008
  [hep-ph/0702003 [hep-ph]].



\bibitem{Kauffman:1991cx}
  R.~P.~Kauffman,
  Phys.\ Rev.\ D {\bf 45} (1992) 1512;
  C.~P.~Yuan,
  Phys.\ Lett.\ B {\bf 283} (1992) 395.
 

\bibitem{Curci:1980uw}
  G.~Curci, W.~Furmanski and R.~Petronzio,
  Nucl.\ Phys.\ B {\bf 175} (1980) 27.


\bibitem{Furmanski:1980cm}
  W.~Furmanski and R.~Petronzio,
  Phys.\ Lett.\ B {\bf 97} (1980) 437.


\bibitem{dimreg}
  Z.~Kunszt, A.~Signer and Z.~Trocsanyi,
  Nucl.\ Phys.\ B {\bf 411} (1994) 397
  [hep-ph/9305239];
  S.~Catani, M.~H.~Seymour and Z.~Trocsanyi,
  Phys.\ Rev.\ D {\bf 55} (1997) 6819
  [hep-ph/9610553];
  Z.~Bern, A.~De Freitas, L.~J.~Dixon and H.~L.~Wong,
  Phys.\ Rev.\ D {\bf 66} (2002) 085002
  [hep-ph/0202271].



\bibitem{ir1loop}
  W.~T.~Giele and E.~W.~N.~Glover,
  Phys.\ Rev.\ D {\bf 46} (1992) 1980;
  Z.~Kunszt, A.~Signer and Z.~Trocsanyi,
  Nucl.\ Phys.\ B {\bf 420} (1994) 550
  [hep-ph/9401294];
  S.~Catani and M.~H.~Seymour,
  Nucl.\ Phys.\ B {\bf 485} (1997) 291
   [Erratum-ibid.\ B {\bf 510} (1998) 503]
  [hep-ph/9605323].

\bibitem{ggFF2}
  R.~V.~Harlander,
  Phys.\ Lett.\ B {\bf 492} (2000) 74
  [hep-ph/0007289];
  V.~Ravindran, J.~Smith and W.~L.~van Neerven,
  Nucl.\ Phys.\ B {\bf 704} (2005) 332
  [hep-ph/0408315].

\bibitem{FF3}
  S.~Moch, J.~A.~M.~Vermaseren and A.~Vogt,
  JHEP {\bf 0508} (2005) 049
  [hep-ph/0507039],
  Phys.\ Lett.\ B {\bf 625} (2005) 245
  [hep-ph/0508055];
  P.~A.~Baikov, K.~G.~Chetyrkin, A.~V.~Smirnov, V.~A.~Smirnov and M.~Steinhauser,
  Phys.\ Rev.\ Lett.\  {\bf 102} (2009) 212002
  [arXiv:0902.3519 [hep-ph]];
  R.~N.~Lee, A.~V.~Smirnov and V.~A.~Smirnov,
  JHEP {\bf 1004} (2010) 020
  [arXiv:1001.2887 [hep-ph]];
  T.~Gehrmann, E.~W.~N.~Glover, T.~Huber, N.~Ikizlerli and C.~Studerus,
  JHEP {\bf 1006} (2010) 094
  [arXiv:1004.3653 [hep-ph]].

\bibitem{3loopsing}
  L.~J.~Dixon, L.~Magnea and G.~F.~Sterman,
  JHEP {\bf 0808} (2008) 022
  [arXiv:0805.3515 [hep-ph]];
  T.~Becher and M.~Neubert,
  JHEP {\bf 0906} (2009) 081
  [arXiv:0903.1126 [hep-ph]].



\bibitem{Catani:1999ss}
  S.~Catani and M.~Grazzini,
  Nucl.\ Phys.\ B {\bf 570} (2000) 287
  [hep-ph/9908523].


\bibitem{Bern:1999ry}
  Z.~Bern, V.~Del Duca, W.~B.~Kilgore and C.~R.~Schmidt,
  Phys.\ Rev.\ D {\bf 60} (1999) 116001
  [hep-ph/9903516].

\bibitem{Catani:2000pi}
  S.~Catani and M.~Grazzini,
  Nucl.\ Phys.\ B {\bf 591} (2000) 435
  [hep-ph/0007142].


\bibitem{Campbell:1997hg} 
  J.~M.~Campbell and E.~W.~N.~Glover,
  Nucl.\ Phys.\ B {\bf 527} (1998) 264
  [hep-ph/9710255].

\bibitem{Kosower:1999rx}
  D.~A.~Kosower and P.~Uwer,
  Nucl.\ Phys.\ B {\bf 563} (1999) 477
  [hep-ph/9903515].



\bibitem{dyampli}
  R.~J.~Gonsalves,
  Phys.\ Rev.\ D {\bf 28} (1983) 1542;
  G.~Kramer and B.~Lampe,
  Z.\ Phys.\ C {\bf 34} (1987) 497
   [Erratum-ibid.\ C {\bf 42} (1989) 504];
  T.~Matsuura and W.~L.~van Neerven,
  Z.\ Phys.\ C {\bf 38} (1988) 623;

\bibitem{Matsuura:1988sm}
  T.~Matsuura, S.~C.~van der Marck and W.~L.~van Neerven,
  Nucl.\ Phys.\ B {\bf 319} (1989) 570.




\bibitem{Catani:2003zt}
  S.~Catani, D.~de Florian, M.~Grazzini and P.~Nason,
  JHEP {\bf 0307} (2003) 028
  [hep-ph/0306211].

\bibitem{Moch:2005ba}
  S.~Moch, J.~A.~M.~Vermaseren and A.~Vogt,
  Nucl.\ Phys.\ B {\bf 726} (2005) 317
  [hep-ph/0506288].

\bibitem{Catani:1990rr}
  S.~Catani, B.~R.~Webber and G.~Marchesini,
  Nucl.\ Phys.\ B {\bf 349} (1991) 635.

\bibitem{Catani:1998tm}
  S.~Catani, M.~L.~Mangano and P.~Nason,
  JHEP {\bf 9807} (1998) 024
  [hep-ph/9806484].

\bibitem{Moch:2004pa}
  S.~Moch, J.~A.~M.~Vermaseren and A.~Vogt,
  Nucl.\ Phys.\ B {\bf 688} (2004) 101
  [hep-ph/0403192],
  Nucl.\ Phys.\ B {\bf 691} (2004) 129
  [hep-ph/0404111].

\bibitem{Vogt:2000ci}
  A.~Vogt,
  Phys.\ Lett.\ B {\bf 497} (2001) 228
  [hep-ph/0010146].

\bibitem{Catani:2001ic}
  S.~Catani, D.~de Florian and M.~Grazzini,
  JHEP {\bf 0105} (2001) 025
  [hep-ph/0102227].


\bibitem{Moch:2005ky}
  S.~Moch and A.~Vogt,
  Phys.\ Lett.\ B {\bf 631} (2005) 48
  [hep-ph/0508265].

\bibitem{Laenen:2005uz}
  E.~Laenen and L.~Magnea,
  Phys.\ Lett.\ B {\bf 632} (2006) 270
  [hep-ph/0508284].

\bibitem{Kramer:1996iq}
  M.~Kramer, E.~Laenen and M.~Spira,
  Nucl.\ Phys.\ B {\bf 511} (1998) 523
  [hep-ph/9611272].

\bibitem{Harlander:2001is}
  R.~V.~Harlander and W.~B.~Kilgore,
  Phys.\ Rev.\ D {\bf 64} (2001) 013015
  [hep-ph/0102241].

\bibitem{Eynck:2003fn}
  T.~O.~Eynck, E.~Laenen and L.~Magnea,
  JHEP {\bf 0306} (2003) 057
  [hep-ph/0305179].

\bibitem{Ravindran:2006cg}
  V.~Ravindran,
  Nucl.\ Phys.\ B {\bf 752} (2006) 173
  [hep-ph/0603041].
  
\bibitem{Grunberg:2006gd}
  G.~Grunberg,
  Phys.\ Rev.\ D {\bf 74} (2006) 111901
  [hep-ph/0609309];
  S.~Friot and G.~Grunberg,
  JHEP {\bf 0709} (2007) 002
  [arXiv:0706.1206 [hep-ph]].

\bibitem{deFlorian:2012za}
  D.~de Florian and J.~Mazzitelli,
  JHEP {\bf 1212} (2012) 088
  [arXiv:1209.0673 [hep-ph]].



\bibitem{Spira:1995rr}
  M.~Spira, A.~Djouadi, D.~Graudenz and P.~M.~Zerwas,
  Nucl.\ Phys.\ B {\bf 453} (1995) 17
  [hep-ph/9504378].
 


\bibitem{Harlander:2005rq}
  R.~Harlander and P.~Kant,
  JHEP {\bf 0512} (2005) 015
[hep-ph/0509189].


\bibitem{Aglietti:2006tp} 
  U.~Aglietti, R.~Bonciani, G.~Degrassi and A.~Vicini,
  JHEP {\bf 0701}, 021 (2007)
  [hep-ph/0611266].
 
\bibitem{Harlander:2009bw}
  R.~V.~Harlander and K.~J.~Ozeren,
  Phys.\ Lett.\ B {\bf 679} (2009) 467
  [arXiv:0907.2997 [hep-ph]];
  A.~Pak, M.~Rogal and M.~Steinhauser,
  Phys.\ Lett.\ B {\bf 679} (2009) 473
  [arXiv:0907.2998 [hep-ph]].


\bibitem{Balazs:1997hv}
  C.~Balazs, E.~L.~Berger, S.~Mrenna and C.~P.~Yuan,
  Phys.\ Rev.\ D {\bf 57} (1998) 6934
  [hep-ph/9712471].
 
\bibitem{Anastasiou:2002zn}
  C.~Anastasiou, E.~W.~N.~Glover and M.~E.~Tejeda-Yeomans,
  Nucl.\ Phys.\ B {\bf 629} (2002) 255
  [hep-ph/0201274].



\end{thebibliography}
\end{document}